\documentclass[floatfix,aps,pre,a4paper,twocolumn,showpacs]{revtex4-1}

\usepackage{graphicx}
\usepackage{bm}

\newcommand{\BE}{\begin{equation}}
\newcommand{\EE}{\end{equation}}

\begin{document}
\title{Elastic waves and transition to elastic turbulence 
in a two-dimensional viscoelastic Kolmogorov flow}

\author{S. Berti$^1$ and G. Boffetta$^{2,3}$} 
\affiliation{$^1$ Laboratoire de Spectrom\'etrie Phyisque, Grenoble, 
UJF-CNRS UMR5588, France\\
$^2$ Dipartimento di Fisica Generale and INFN, Universit\`a di Torino, 
via P. Giuria 1, 10125 Torino, Italy\\
$^3$ CNR-ISAC, Sezione di Torino, corso Fiume 4, 10133 Torino, Italy}

\date{\today}
\begin{abstract}
We investigate the dynamics of the two-dimensional periodic Kolmogorov flow of
a viscoelastic fluid, described by the Oldroyd-B model, by means of direct
numerical simulations. Above a critical Weissenberg number the flow displays a
transition from stationary to randomly fluctuating states, via periodic ones. The
increasing complexity of the flow in both time and space at progressively
higher values of elasticity accompanies the establishment of mixing features.
The peculiar dynamical behavior observed in the simulations is found to be
related to the appearance of filamental propagating patterns, which develop
even in the limit of very small inertial non-linearities, thanks to the
feedback of elastic forces on the flow.
\end{abstract}
\maketitle
\section{Introduction}
\label{sec:sec01}
It is well known that the addition of small amounts of long chain 
polymers can have a strong impact on flowing fluids' properties. 
One of the most remarkable effects of highly viscous 
polymer solutions which has been recently observed in experiments 
is the development of an ``elastic turbulence'' regime in the limit of
small Reynolds number and strong elasticity \cite{GS00}. 
The flow in this regime displays a rich dynamics, of the type usually 
encountered in high Reynolds number turbulent flow in Newtonian fluids. 
Typical features of turbulent flows (e.g., broad range of active scales, 
growth of flow resistance) are observed in the experiments by 
Steinberg and collaborators~\cite{GS00,BSS07}, even at low velocity and 
high viscosity, i.e. in the limit of vanishing Reynolds number. 
As a consequence of turbulent-like motion, elastic turbulence has been
proposed as an efficient technique for mixing in very low Reynolds
flows, such as in microchannel flows \cite{GS01,BSBGS04,AVG05}.
Despite its great technological interest, elastic turbulence
is still only partially understood from a theoretical point of view.
Presently available theoretical predictions are based on simplified versions of
viscoelastic models and on the analogy with magneto-hydrodynamics (MHD) equations \cite{BFL01,FL03}. 
In a previous work~\cite{BBBCM08} we showed, by means of numerical simulations,
that the basic phenomenology of elastic turbulence is reproduced by a 
simple model of polymeric fluid and in an idealized geometry, namely 
a two-dimensional periodic shear flow without boundaries. 
Notwithstanding quantitative differences, most observed features have 
a strong qualitative resemblance with experimental results. 

The mechanism through which polymer molecules can influence
properties of a flow is their extreme extensibility. 
Polymers, typically composed by a large number of monomers,
at equilibrium are coiled in a ball of radius $R_0$.
In presence of a non-homogeneous flow, the molecule is deformed
in an elongated structure characterized by its
end-to-end distance $R$ which can be much larger than $R_0$.
The deformation of molecules is the result of the competition
between the stretching induced by velocity gradients and
the entropic relaxation of polymers to their equilibrium configuration.
Experiments with DNA molecules \cite{PSC94,QBC97} show that
this relaxation is linear, provided that the elongation is
small compared with the maximal extension $R \ll R_{max}$,
and therefore can be characterized by a typical relaxation time 
$\tau$ \cite{H77}. 
Besides the Reynolds number $Re$, commonly used in hydrodynamics, 
in the study of flowing polymer solutions, it is useful to introduce 
the Weissenberg number $Wi$. This is defined as the product of $\tau$ 
and the characteristic deformation rate, and it 
measures the  relative importance between polymer relaxation and stretching.
When $Wi \ll 1$ relaxation is fast compared to the stretching time
and polymers remain in the coiled state. For $Wi \gg 1$, on the
contrary, polymers are substantially elongated. The transition point
is called the coil-stretch transition and occurs at $Wi=O(1)$. 
For Weissenberg numbers much larger than unity, 
polymers can provide a significant feedback effect on the dynamical 
behavior of the fluid they are suspended into, 
so that they behave as active objects capable of modifying the properties 
of the flow.

Because of the feedback effect, above the coil-stretch transition, 
a laminar flow can become unstable even in the small $Re$ limit 
of vanishing inertial non-linearities. Purely elastic instabilities 
have been studied in different flow configurations. 
In a curvilinear shear flow as, e.g., Couette flow between rotating 
cylinders, the driving force of the elastic instability is the 
``hoop stress''~(see, for instance, \cite{BCAH87} page~62), a volume force in the direction of 
curvature due to the first normal stress difference. 
The mechanism for this instability was first proposed in~\cite{LSM90} 
and experimentally verified in~\cite{GS98}. 
In the framework of flows with curvilinear streamlines, a dimensionless 
criterion for the critical onset conditions, which applies to various 
flow geometries, was established in~\cite{PMK96,MKPO96}.
The role of flow-microstructure coupling in pattern formation in the 
Taylor-Couette configuration was also examined in numerical simulations of 
the $3D$ FENE-P (finitely extensible non-linear elastic-Peterlin) 
model~\cite{TSK06}. 
On the other hand, purely elastic instabilities in a parallel flow with 
rectilinear streamlines were obtained for the first time in a $2D$ Kolmogorov 
flow of a viscoelastic fluid described by the linear Oldroyd-B 
model~\cite{BCMPV05}. Also for a more realistic shear flow of a viscoelastic 
fluid in a straight channel, there exist theoretical results exhibiting 
the elastic instability~\cite{MS07}, although they are still not 
experimentally verified. 
Elastically induced pattern formation and production of fluid mixing were 
also studied in extensional flows, in experiments~\cite{ATDG06}, and 
numerical simulations of a Stokesian linear viscoelastic model~\cite{TS09}.

In this paper, we study the dynamics of dilute polymer solutions by means
of direct numerical simulations of a general linear viscoelastic model, in 
a flow configuration corresponding to the two-dimensional Kolmogorov shear 
flow. Our simulations for values of the stability parameters ($Re$ and $Wi$) 
close to the elastic instability threshold, exhibit a transition to 
new intriguing flow states, in the form of ``elastic waves'', driving the 
establishment of mixing features. 
In particular, at growing the elasticity of the solution, after a first 
transition to steady uniformly traveling patterns with different symmetry 
with respect to the reference laminar fixed point, we detect 
the appearance of temporal oscillations and, eventually, the onset of a 
strongly non-linear multiscale state disordered in both space and time.

The article is organized as follows: in the next section we introduce the 
viscoelastic model. In section~\ref{sec:sec03a} we present numerical 
results concerning the transition to mixing states in terms of an analysis 
of the behavior of global observables. 
In section~\ref{sec:sec03b} we show the results of numerical measurements 
concerning the spatiotemporal structure of the patterns emerging above 
the onset of the instability. Finally, section~\ref{sec:sec04} summarizes 
our conclusions.

\section{Model}
\label{sec:sec02}
To describe the dynamics of a dilute polymer solution we
adopt the well known linear Oldroyd-B model \cite{BCAH87}
\BE
\partial_t \bm{u} + (\bm{u} \cdot \bm{\nabla}) \bm{u} = - \bm{\nabla} p + \nu \Delta \bm{u} 
+ \frac{2 \eta \nu}{\tau} \bm{\nabla}  \cdot \bm{\sigma}  +\bm{f}, 
\label{eq:eq01} 
\EE
\BE
\partial_t \bm{\sigma} + (\bm{u} \cdot \bm{\nabla})\bm{\sigma} = (\bm{\nabla} \bm{u})^T \cdot \bm{\sigma}
+ \bm{\sigma} \cdot (\bm{\nabla} \bm{u}) - 2\frac{( \bm{\sigma}-\bm{1})}{\tau},
\label{eq:eq02}
\EE
where ${\bm u}$ is the incompressible velocity field, ${\bm \sigma}$
the symmetric positive definite matrix representing
the normalized conformation tensor of polymer molecules and ${\bm 1}$
is the unit tensor. 
The trace $tr{\bm \sigma}$ is a measure of local polymer elongation.
The solvent viscosity is denoted by $\nu$ and
$\eta$ is the zero-shear contribution of polymers to the total solution
viscosity $\nu_{t}=\nu(1+\eta)$ and is proportional to
the polymer concentration.
In absence of flow, ${\bm u}=0$, polymers relax to the equilibrium
configuration and ${\bm \sigma}={\bm 1}$. 
The extra stress term 
$\frac{2 \eta \nu}{\tau} \bm{\nabla}  \cdot \bm{\sigma}$ 
in (\ref{eq:eq01}) takes into account the elastic forces of 
polymers and provides a feedback mechanism on the flow. 

In order to concentrate on the intrinsic dynamics of the viscoelastic solution
we use the simple geometrical setup of the periodic Kolmogorov flow in 
two dimensions \cite{AM60}.
With the forcing ${\bm f}=(F\cos(y/L),0)$, the system of
equations (\ref{eq:eq01}-\ref{eq:eq02}) has a laminar Kolmogorov fixed
point given by
\begin{eqnarray}
& & {\bm u}_0=(U_0\cos(y/L),0) \nonumber \\
& &  \label{eq:eq03} \\
& & {\bm \sigma}_0=
\left(
\begin{array}{cc}
1+{\tau}^2\, {U_0^2 \over 2 L^2}\sin^2{(y/L)} \,\,\, & 
\,\,\, -{\tau} {U_0 \over 2 L} \sin{(y/L)} \\\\
-{\tau} {U_0 \over 2 L} \sin{(y/L)} \,\,\, & \,\,\,1
\end{array}
\right) \nonumber
\end{eqnarray}
with $F=[\nu U_0 (1+\eta)]/L^2$ \cite{BCMPV05}.
The laminar flow fixes a characteristic
scale $L$, velocity $U_0$ and time $T=L/U_0$. In terms of these variables,
one can define the Reynolds number as $Re={U_0L \over \nu_t}$ and the
Weissenberg number as $Wi={\tau U_0 \over L}$. The ratio of these numbers
defines the elasticity of the flow $El=Wi/Re$. Notice these are a-priori 
estimates of the non-dimensional parameters; in the following we will 
refer to them as $Re_0$ and $Wi_0$. In the presence of velocity fluctuations 
superimposed to a non-zero mean flow $U$, it is also possible to compute 
a-posteriori non-dimensional parameters as 
$Re={UL \over \nu_t}$ and $Wi={\tau U \over L}$, once the average 
velocity $U$ has been computed; of course, in general $U \neq U_0$.

It is well known that the Kolmogorov flow displays instability with
respect to large-scale perturbations, i.e. with wavelength much
larger than $L$. In the Newtonian case, the instability arises
at $Re_{c}=\sqrt{2}$ \cite{MS61}.
The presence of polymers changes the stability diagram of laminar 
flows \cite{Larson92,GS96} and can induce elastic instabilities
which are not present in the Newtonian limit 
\cite{LSM90,Shaqfeh96,BCMPV05,GS98},
even in the case of periodic flows \cite{ASK02}.
In this respect, the Kolmogorov flow is no exception, and recent
analytical and numerical investigations have offered the complete instability
diagram in the ($Wi$, $Re$) plane \cite{BCMPV05}.
For the purpose of the present work, we just have to recall
that linear stability
analysis shows that for sufficiently large values of elasticity,
the Kolmogorov flow displays purely elastic instabilities, even at
vanishing Reynolds number (see Fig.~1 of \cite{BCMPV05}).
Note that in this case the direction of the most unstable mode is 
at a small angle with respect to the basic flow, at variance with 
the purely hydrodynamic transverse instabilities. In the Newtonian
case, non-transverse instabilities are observed for less symmetric basic 
flows only \cite{df_pra91}.
We also remark that the fact that the basic flow has rectilinear
streamlines does not exclude the onset of the elastic 
instability \cite{MS07}.
Above the elastic instability the flow can develop a disordered secondary flow
which persists in the limit of vanishing Reynolds number and eventually
leads to the elastic turbulence regime \cite{GS04}.

\section{Analysis and results}
\label{sec:sec03}
The equations of motion (\ref{eq:eq01})-(\ref{eq:eq02}) are integrated by
means of a pseudo-spectral method implemented on a two-dimensional grid
of size $L_0=2 \pi$ with periodic boundary conditions at resolution
$512^2$.
It is well known that the integration of viscoelastic models is 
limited by numerical instabilities associated with the loss
of positiveness of the conformation tensor \cite{SB95}.
These instabilities are particularly important at high elasticity
and limit the possibility to investigate the elastic turbulent regime
by direct implementation of equations (\ref{eq:eq01})-(\ref{eq:eq02}).
To overcome this problem, we have implemented an algorithm based on a
Cholesky decomposition of the conformation matrix that ensures
symmetry and positive definiteness \cite{VC03}.
While this allows us to safely reach high elasticity it has a
price to pay in terms of limited resolution and large computing time. 
In the following, we will present the numerical results obtained using
this approach. 
\subsection{Transition to mixing states}
\label{sec:sec03a}
In order to investigate the transition to mixing states, we chose to 
work in a region of parameters close to the onset of the purely elastic 
instability. Here we fix $Re=1$ and increase the Weissenberg number from 
$Wi=6$ to $Wi \gtrsim 20$. 
The initial condition is obtained by adding a small random perturbation to 
the fixed point solution (\ref{eq:eq03}) and the system is evolved in time 
until a (statistically) steady state is reached. 

We start the analysis by considering the behavior of global quantities, 
such as the kinetic energy 
$K\equiv\langle |\bm{u}|^2 \rangle/2$ and the average square polymer elongation 
$\Sigma\equiv\langle tr \bm{\sigma} \rangle$.
The total energy of the system $E_T=K+{{2\nu\eta} \over \tau^2}\Sigma$ is the sum of kinetic and elastic contributions, 
and its balance equation, from~(\ref{eq:eq01})-(\ref{eq:eq02}) is:
\BE
{{d E_T} \over {dt}} = \epsilon_I-\epsilon_\nu -{{2\nu\eta} \over \tau^2}\left(\Sigma-2\right)
\label{eq:eq06}
\EE
where $\epsilon_I=\langle\bm{f} \cdot \bm{u}\rangle$ is the energy input, 
$\epsilon_\nu=\nu\langle \left(\partial_\alpha u_\beta\right)^2\rangle$ 
is the viscous dissipation, and the last term represents elastic dissipation. 
The balance of forcing and dissipation produces 
a steady state, ${{dE_T} \over {dt}}=0$ in which the total energy is 
constant, at least in a statistical sense (i.e. on time scales larger than 
the typical duration of fluctuations).

\begin{figure}[!h]
\includegraphics[draft=false,scale=0.6]{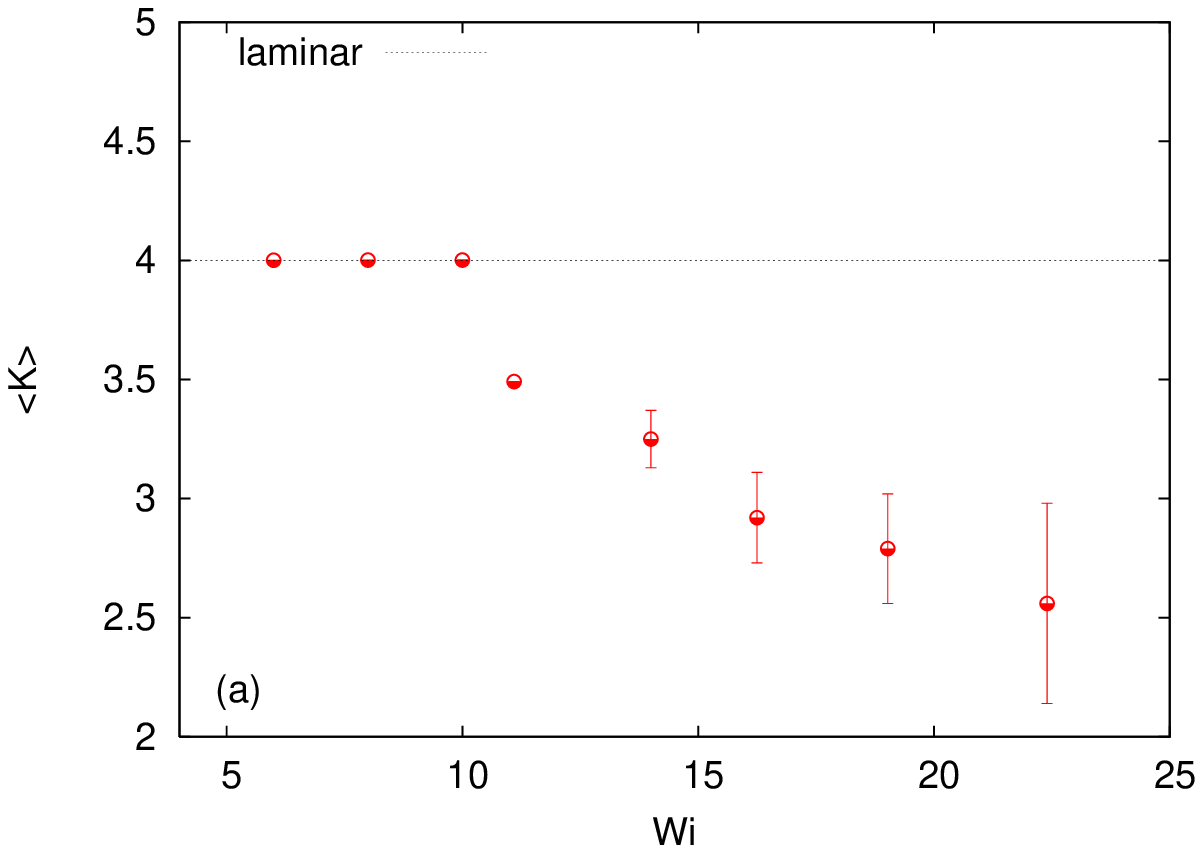}
\includegraphics[draft=false,scale=0.6]{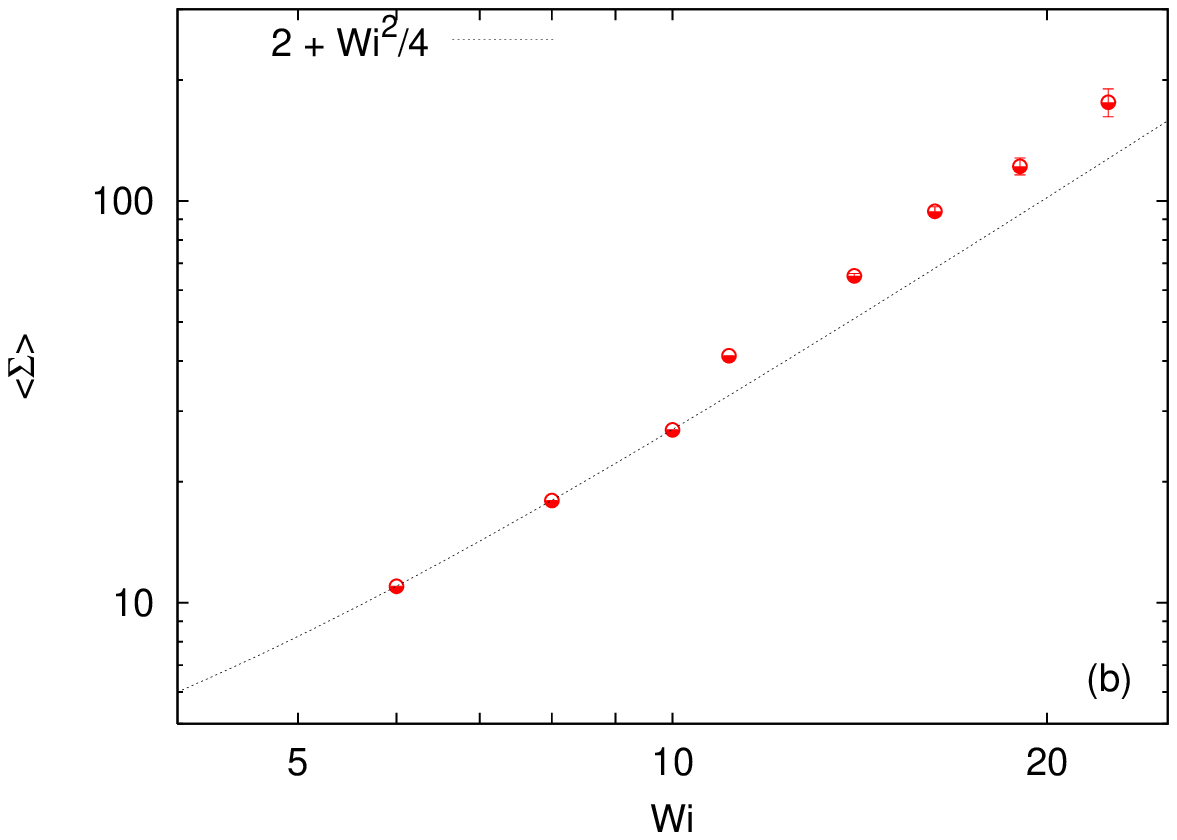}
\caption{(Color online) Mean kinetic energy $\langle K \rangle$ (a) and 
mean square polymer elongation $\langle \Sigma \rangle$ (b) in statistically
stationary states as a function of $Wi$ for $Re\simeq1$. 
Dashed lines are the laminar predictions from (\ref{eq:eq03}) and error bars 
represent the amplitude of fluctuations.}
\label{fig:fig01}
\end{figure}

The mean values of $K$ and $\Sigma$ as a function of 
the Weissenberg number are presented in Fig.~\ref{fig:fig01}. 
Here and in the following the Weissenberg number $Wi=U\tau/L$ and the 
Reynolds number $Re=UL/\nu_t$ are defined in terms of the amplitude $U$ 
of the mean velocity, which is measured from velocity profiles. 
Indeed, a remarkable feature of the Kolmogorov flow 
is that even in the turbulent regime the mean velocity and conformation 
tensor are accurately described by sinusoidal profiles \cite{BCM05}, though 
with different amplitudes with respect to the laminar fixed point. 

As it is shown in Fig.~\ref{fig:fig01}, the system does not depart from 
the laminar fixed point up to $Wi=10$ which is approximately the 
critical $Wi$ obtained in \cite{BCMPV05}. 
For higher values of the Weissenberg number, a depletion in the kinetic 
energy is observed, accompanied by a growth in the trace of the conformation 
tensor, larger than its laminar prediction 
$\langle tr \bm{\sigma} \rangle=2+Wi^2/4$. 
Therefore, polymers drain energy from the mean flow in order to 
elongate and, once sufficiently stretched they appreciably start to 
back-react on the flow itself, through the coupling term 
${{2 \eta \nu} \over \tau} \bm{\nabla} \cdot \bm{\sigma}$ in 
(\ref{eq:eq02}). 
The increase with respect to the laminar growth indicates this coupling is very efficient in 
sustaining the stretching of polymers.

\begin{figure}[!h]
\includegraphics[draft=false,scale=0.65]{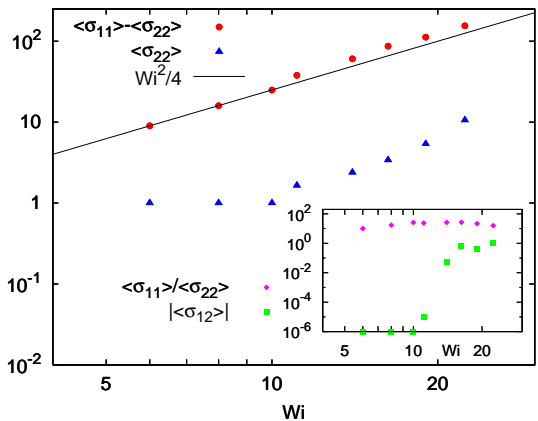}
\caption{(Color online) Average values of $\langle \sigma_{11} \rangle -
\langle \sigma_{22} \rangle$, 
proportional to the first normal stress difference $N_1$, and of
$\langle \sigma_{22} \rangle$ as a function of $Wi$. The line corresponds 
to the $Wi^2$ behavior of the laminar fixed point. Inset: ratio 
$\langle \sigma_{11} \rangle/\langle \sigma_{22} \rangle$ and
$|\langle \sigma_{12} \rangle|$ as a function of $Wi$.}
\label{fig:fig02}
\end{figure}

We observe here that, due to the present flow geometry, one of the 
diagonal elements of the conformation tensor is much larger than the 
other, namely $\sigma_{11} \gg \sigma_{22}$. This implies that the 
trace of the tensor is almost entirely given by $\sigma_{11}$, and 
the same is true for the first normal stress difference 
$N_1 \propto (\sigma_{11}-\sigma_{22}) \approx \sigma_{11}$, 
so that $\langle tr \bm{\sigma} \rangle \approx \langle N_1 \rangle $. 
Together with the growth of the trace of the conformation tensor, 
the growth of the first normal stress difference is a clear 
signature of the elasticity of the flow. 
An inspection at the behavior of the components of the conformation 
tensor, Fig.~\ref{fig:fig02}, indeed reveals that 
$\langle \sigma_{11} \rangle \gg \langle \sigma_{22} \rangle$ and, in turn, 
$\langle \sigma_{22} \rangle \gg |\langle \sigma_{12} \rangle|$.
In particular we find 
$\langle \sigma_{11} \rangle \approx 10 \langle \sigma_{22} \rangle$ 
independently of $Wi$, and 
$\langle \sigma_{22} \rangle \approx 10 |\langle \sigma_{12} \rangle|$ 
for $Wi \gtrsim 10$. 
Although $\langle \sigma_{12} \rangle$ and $\langle \sigma_{22} \rangle$ 
always remain small compared to $\langle \sigma_{11} \rangle$, 
their behavior with Weissenberg number clearly shows a transition around 
$Wi \approx 10$, indicating the establishment of a nonzero transversal 
force, growing with $Wi$, potentially capable of sustaining a secondary 
flow with nonzero component in the direction perpendicular to that of the 
base flow.

\begin{figure}[!h]
\includegraphics[draft=false,scale=0.6]{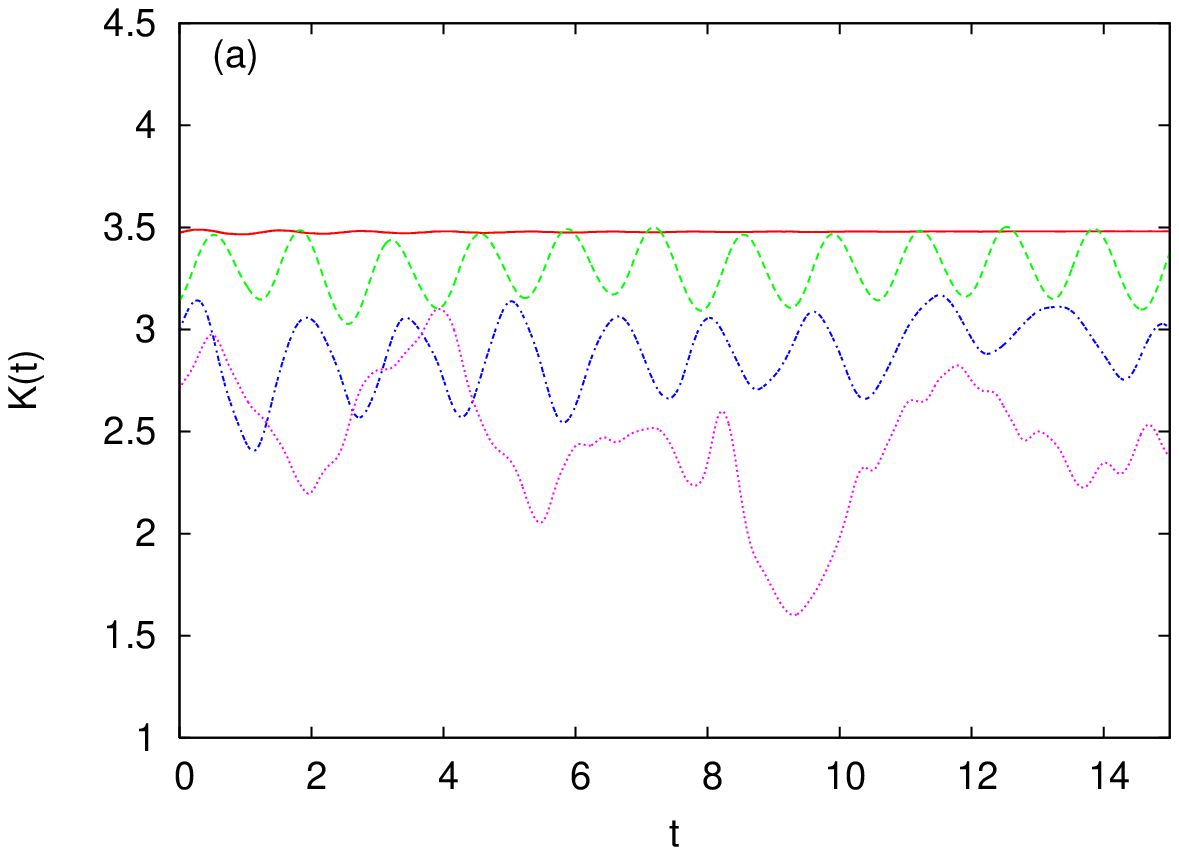}
\includegraphics[draft=false,scale=0.6]{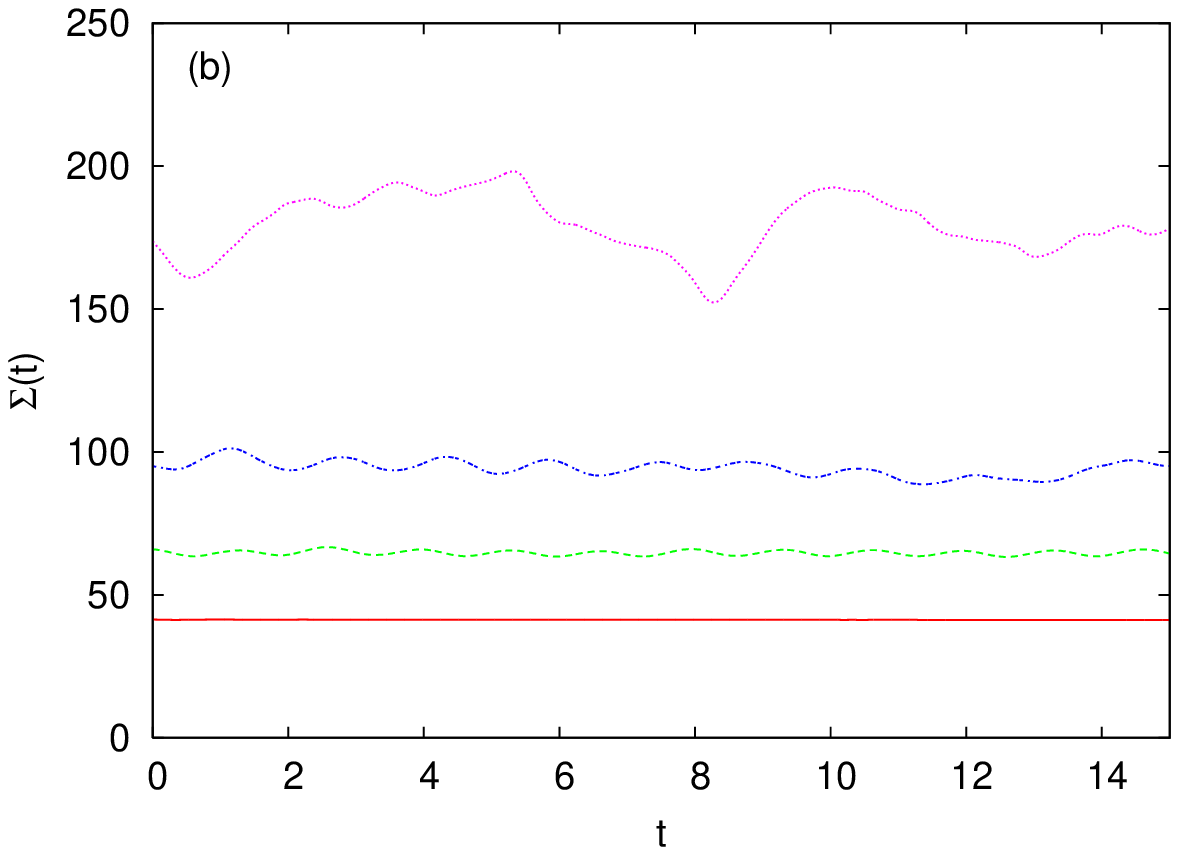}
\caption{(Color online) Time behavior of kinetic energy $K$ (a) and 
average square polymer elongation $\Sigma$ (b) in statistically
stationary states, for $Re\simeq1$ at $Wi=11.1$ (continuous red lines), 
$Wi=14$ (dashed green lines), $Wi=16.25$ (dot-dashed blue lines) and $Wi=22.4$ (dotted pink lines).
In (a) $Wi$ increases from top to bottom, while in (b) from bottom to top,
in agreement with the mean values shown in Fig.~\ref{fig:fig01}.}
\label{fig:fig03}
\end{figure}

Let us now discuss the behavior of temporal fluctuations of the global 
quantities $K$ (kinetic energy) and $\Sigma$ 
(trace of the conformation tensor), when the Weissenberg number is 
increased, which provides further information on the change 
of dynamical regime. In Fig.~\ref{fig:fig03} we report $K$ and 
$\Sigma$ versus time, for several values of $Wi$, above the elastic 
instability threshold $Wi_c \approx 10$.
The picture emerging from this figure is that of a transition from 
constant to randomly fluctuating states, via periodic ones.

Below $Wi=14$, $K$ and $\Sigma$ are constant, although at different values 
with respect to the laminar fixed point. 
We remark that the constant behavior does not necessarily mean the 
flow is time independent, as it is also compatible with the presence of 
steady, uniformly traveling patterns of the form $A(x-vt)$ propagating
with velocity $v$.

\begin{figure}[!h]
\includegraphics[draft=false,scale=0.45]{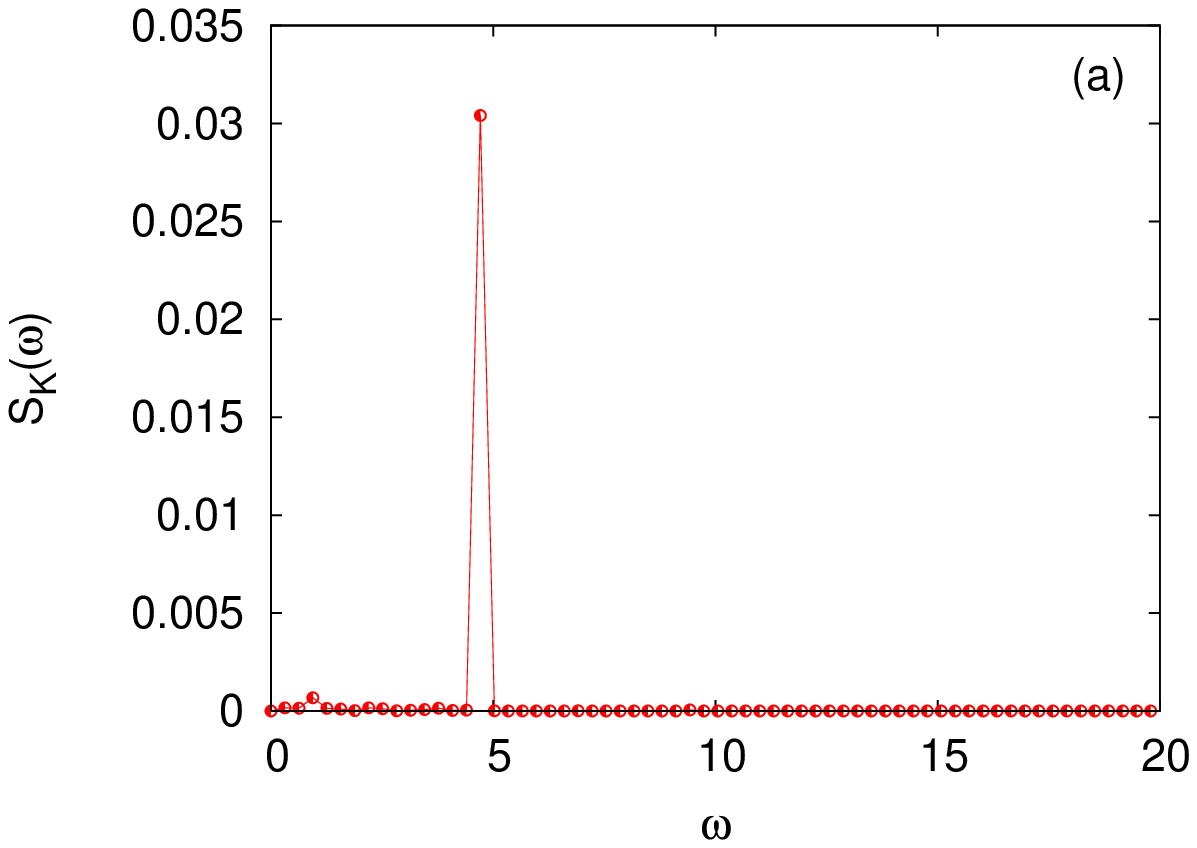}
\includegraphics[draft=false,scale=0.45]{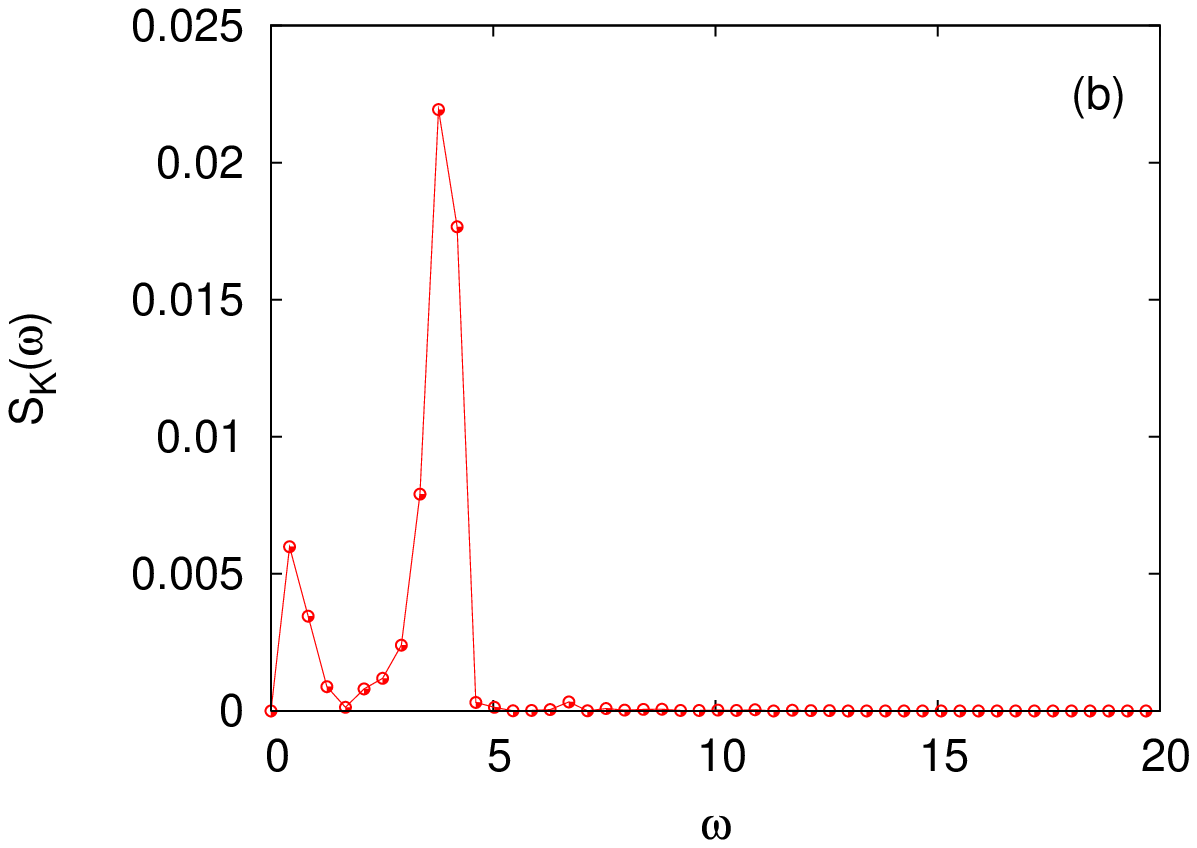}
\includegraphics[draft=false,scale=0.45]{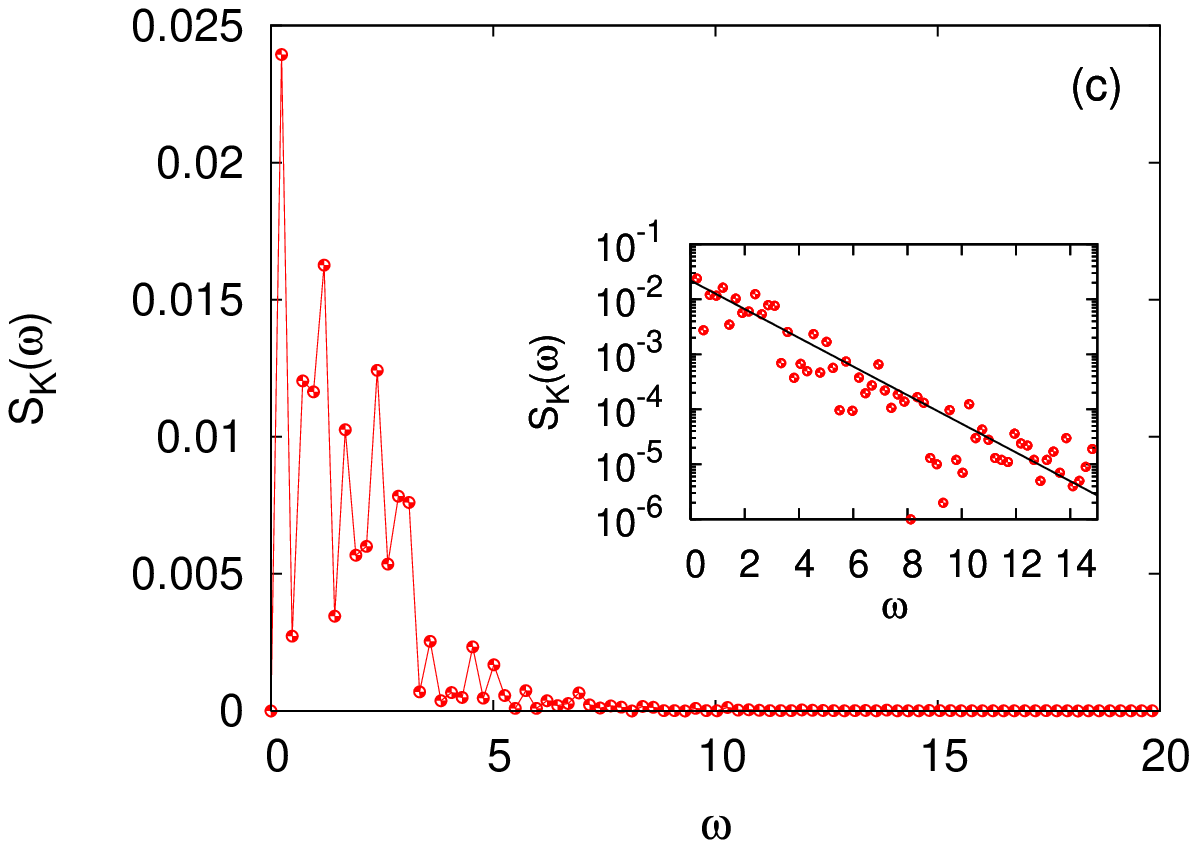}
\caption{(Color online) Frequency spectra $S_K(\omega)$ of $K(t)$ at 
$Re=0.875$, $Wi=14$ (a); $Re=0.8125$, $Wi=16.25$(b);
$Re=0.7925$, $Wi=19.02$ (c). The inset in (c) is the frequency spectrum 
plotted with lin-log axes to show the exponential decay (black line).}
\label{fig:fig04}
\end{figure}

When $Wi \geq 14$ the kinetic energy and the squared polymer elongation 
oscillate around their average values. 
To assess the frequency content of the above fluctuating temporal signals, 
we performed a Fourier analysis, computing the power spectrum
\BE
S_K(\omega)=\left|{1 \over T_{\mbox{\scriptsize max}}} 
\int_0^{T_{\mbox{\scriptsize max}}} K(t) e^{i \omega t}\right|^2
\label{eq:eq07}
\EE
where $[0,T_{\mbox{\scriptsize max}}]$ is an interval of time in 
the statistically steady state; the results are shown in 
Fig.~\ref{fig:fig04} (similar results are obtained for $\Sigma$).
This procedure allowed to identify two transitions: at $Wi \geq 14$ an
explicit time dependency appears, signaled by a single frequency peak 
in the spectra of $K$ and $\Sigma$. 
In this range of parameters the system is in a periodic state, 
characterized by global oscillations in time. 
The peak frequency $\omega_*$ is found to be slightly larger than the 
frequency $\omega_A=2\pi/T_A$ associated to 
the advective time scale, $T_A=L_0/U$, and smaller than that associated to 
the polymer relaxation time $\omega_\tau=2\pi/\tau$ 
(which is in turn much smaller than the frequency 
$\omega_\Gamma=2\pi/T_\Gamma$ related to the inverse 
velocity gradient $T_\Gamma=L/U$ of the mean flow): 
$\omega_A \lesssim \omega_* < \omega_\tau \ll \omega_\Gamma$. 
The time dependency of the kinetic energy implies that of the velocity 
field $\bm{u}=\bm{u}(\bm{x},t)$. 
It is known from the theory of dynamical systems that this is a 
sufficient condition for the appearance of Lagrangian 
chaos in a $2D$ incompressible flow and
the chaoticity of Lagrangian trajectories implies the establishment of 
mixing features \cite{CCV10}.
In previous numerical simulations \cite{BBBCM08}, we measured the Lagrangian 
Lyapunov exponent $\lambda$, defined as the mean rate of separation of 
two infinitesimally close particles transported by the flow. 
This quantity, which gives an estimation of the inverse mixing time, was 
indeed found to be positive in a corresponding range of $Wi \gtrsim 10$ 
values, implying the onset of chaotic (Lagrangian) dynamics. 

The present picture persists at increasing Weissenberg numbers,  until
at $Wi=16.25$ a second frequency, associated to a slower time scale, shows up.
For $Wi \geq 19.02$ several frequencies are present, indicating more 
disordered oscillations. In this regime, the spectral amplitudes seem 
to approximately follow an exponential decay.

\begin{figure}[!h]
\includegraphics[draft=false,scale=0.27]{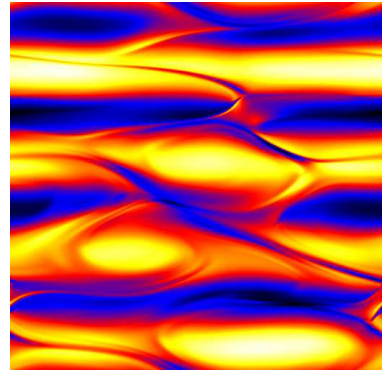} 
\caption{(Color online) Snapshot of the vorticity field 
$\zeta(x,y) = \nabla \times {\bf u}(x,y)$
for $Re \simeq 0.7$ and $Wi \simeq 22.4$. Black (white) corresponds 
to minimum (maximum) vorticity.}
\label{fig:fig05}
\end{figure}

In this latter case, the random fluctuations in time of the global 
quantities point at the onset of irregular 
dynamics associated to a chaotic, mixing, multiscale state. A fixed 
time snapshot of the vorticity field 
$\zeta(x,y)=\partial_x u_y - \partial_y u_x$ at $Wi=22.4$ is 
shown in Fig.~\ref{fig:fig05} where an irregular, disordered pattern
is clearly observable superimposed to the basic periodic flow.

\begin{figure}[!h]
\includegraphics[draft=false,scale=0.6]{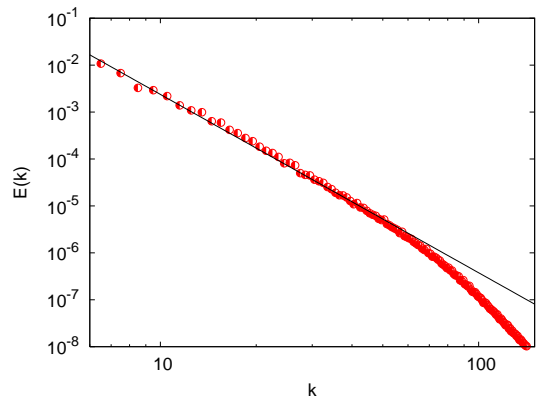}
\caption{(Color online) Spectrum of velocity fluctuations $E(k)$, 
in the wavenumber domain for $Re \simeq 0.7$ and $Wi \simeq 22.4$; 
the line corresponds to the best fit $k^{-3.8}$. 
The spectrum is computed by averaging over several realizations
of the velocity field.}
\label{fig:fig06}
\end{figure}

The statistical characterization of this turbulent-like
state is given by the spectrum of velocity fluctuations in the 
wavenumber $k$ domain. 
As shown in Fig.~\ref{fig:fig06}, the spectrum displays a power-law behavior 
$E(k)\sim k^{-\alpha}$, with exponent $\alpha > 3$, reflecting the 
activity of a whole interval of spatial modes. 
The fast decay of the spectrum, tells us that the velocity and velocity 
gradients are primarily determined by the integral scale, i.e. on
the system size. This implies that elastic turbulence corresponds to a 
temporally random, spatially smooth flow, dominated by strongly 
non-linear interactions between few large scale modes. 

\subsection{Wavy patterns}
\label{sec:sec03b}
As discussed in the previous section, for values of the Weissenberg number
larger than $Wi_c\approx10$, we observe a transition to non stationary states.
We now focus on features of such states in space and in time.  

\begin{figure}[!h]
\begin{center}
\includegraphics[draft=false,scale=0.22]{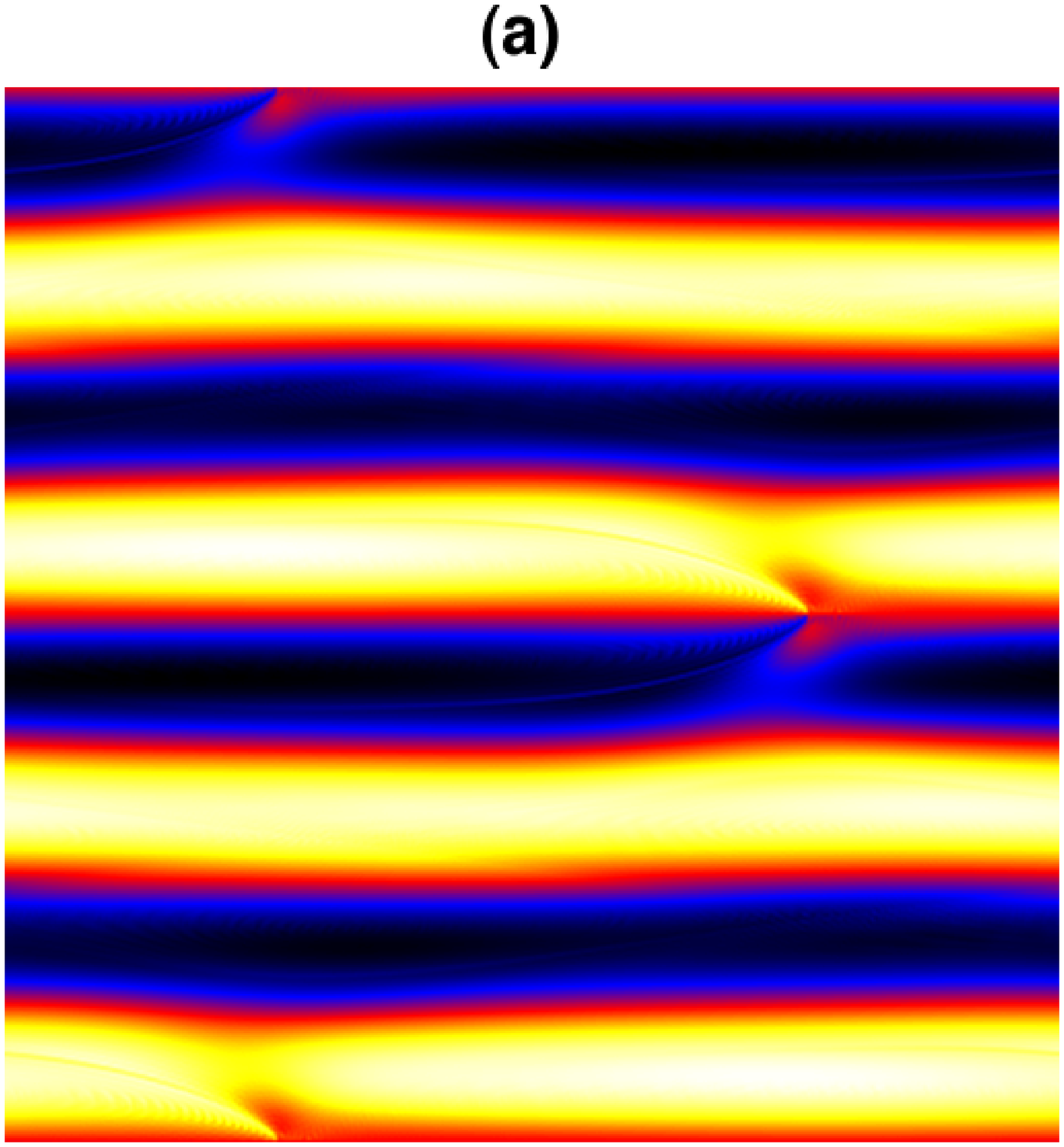} 
\includegraphics[draft=false,scale=0.22]{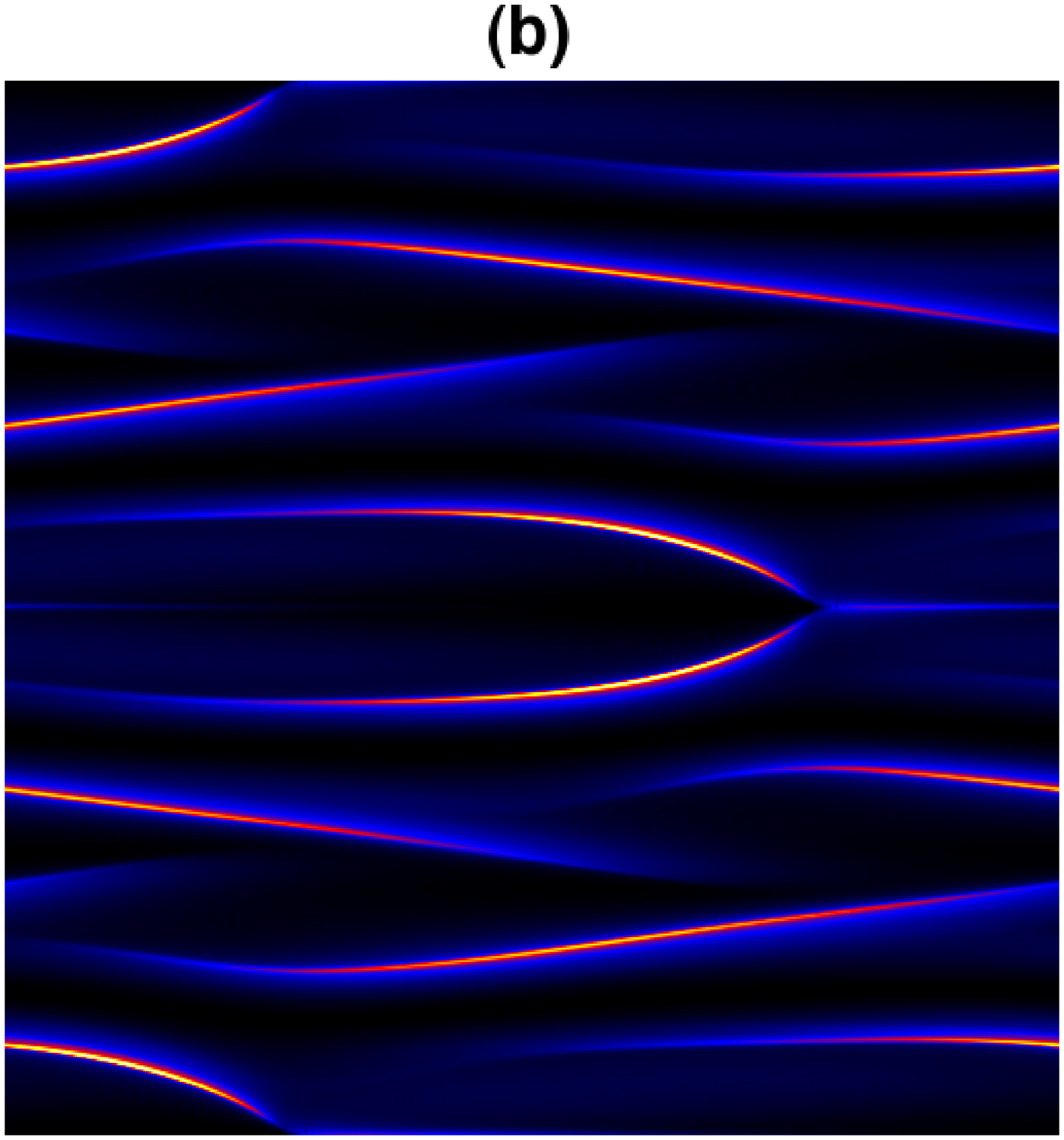}
\includegraphics[draft=false,scale=0.22]{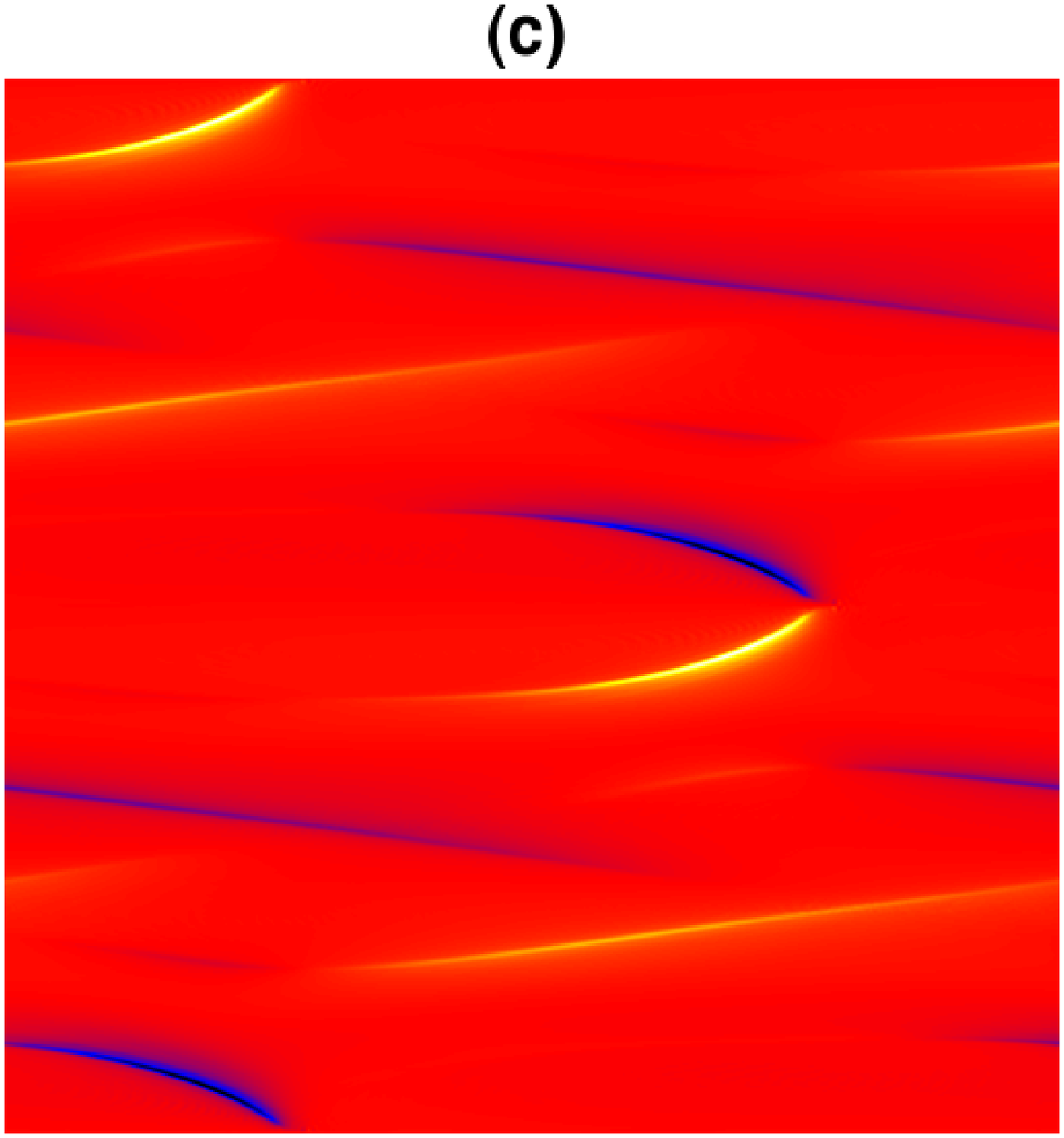} 
\includegraphics[draft=false,scale=0.22]{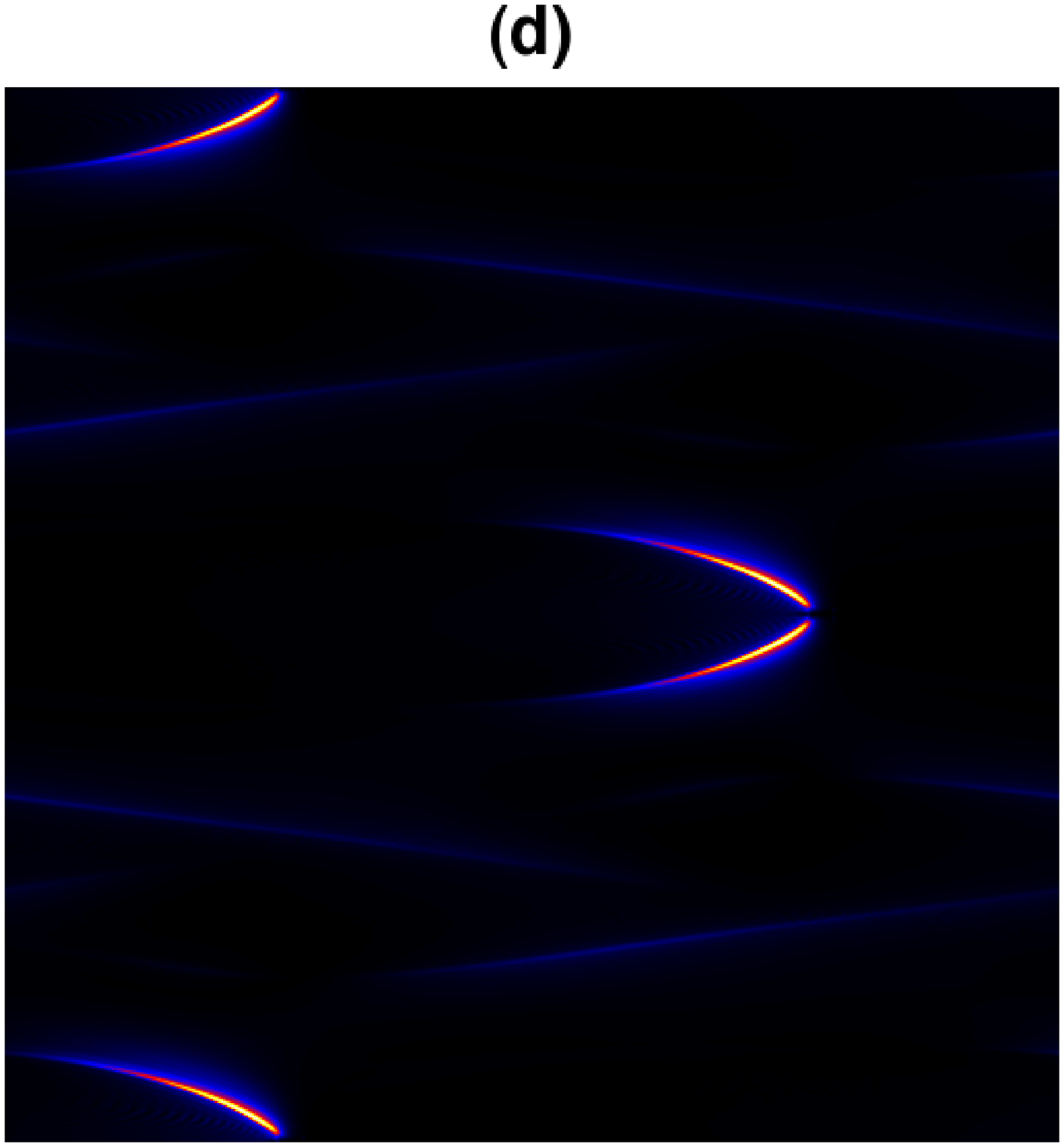}
\end{center}
\caption{(Color online) Snapshots of different fields at a given time: (a) 
$\zeta(x,y)$, (b) $\sigma_{11}(x,y)$, (c) $\sigma_{12}(x,y)$, (d) 
$\sigma_{22}(x,y)$; $Re=0.925$, $Wi=11.1$. 
In all cases black means minimum and white maximum intensity.}
\label{fig:fig07}
\end{figure}

For $Wi \simeq 11$ a typical filamental pattern emerges in both velocity (or
vorticity) and $\sigma$ fields.  Snapshots of these fields in the $(x,y)$ plane
are shown in Fig.~\ref{fig:fig07}. These filamental structures are found
to perform a wavy motion along the horizontal direction of the mean flow. The
possibility of observing ``elastic waves'' in polymer solutions was
theoretically predicted within a simplified uniaxial elastic model \cite{FL03}
which has strong formal analogies with MHD equations, but their experimental
observation is still lacking.

In a narrow range of $Wi$ the patterns are found to simply translate without
changing their form. Already at $Wi=14$ temporal oscillations of pattern shapes
are manifest, particularly on filament tails and in regions between them.
Nevertheless, wavy motion is still clearly detectable. At larger Weissenberg
number, filamental patterns are significantly less stable. Indeed they are
observed to form and disrupt in the course of time, as they travel in both
positive and negative horizontal directions.  

\begin{figure}[!h]
\begin{center}
\includegraphics[draft=false,scale=0.3]{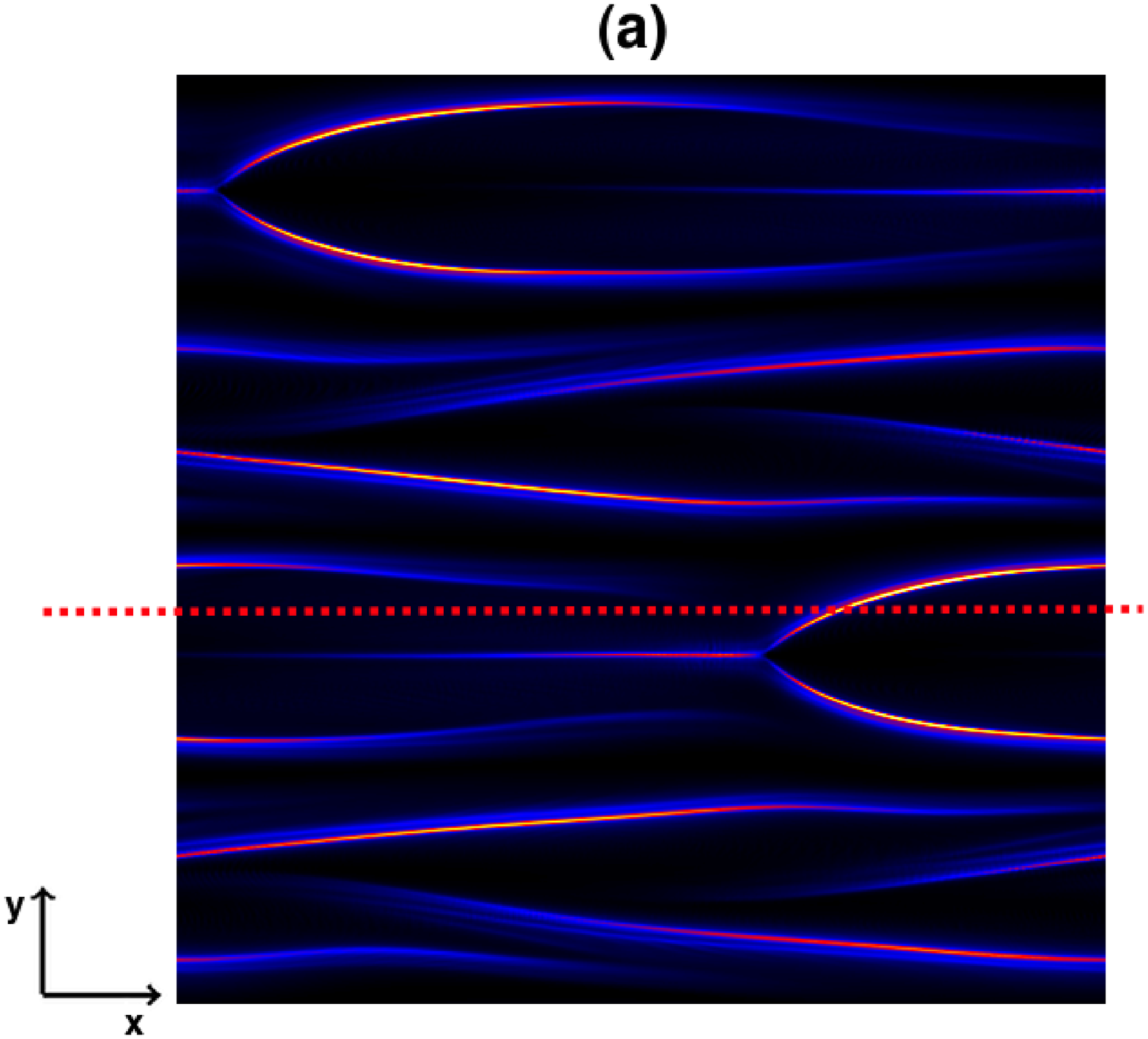} 
\hspace{0.05truecm}
\includegraphics[draft=false,scale=0.3]{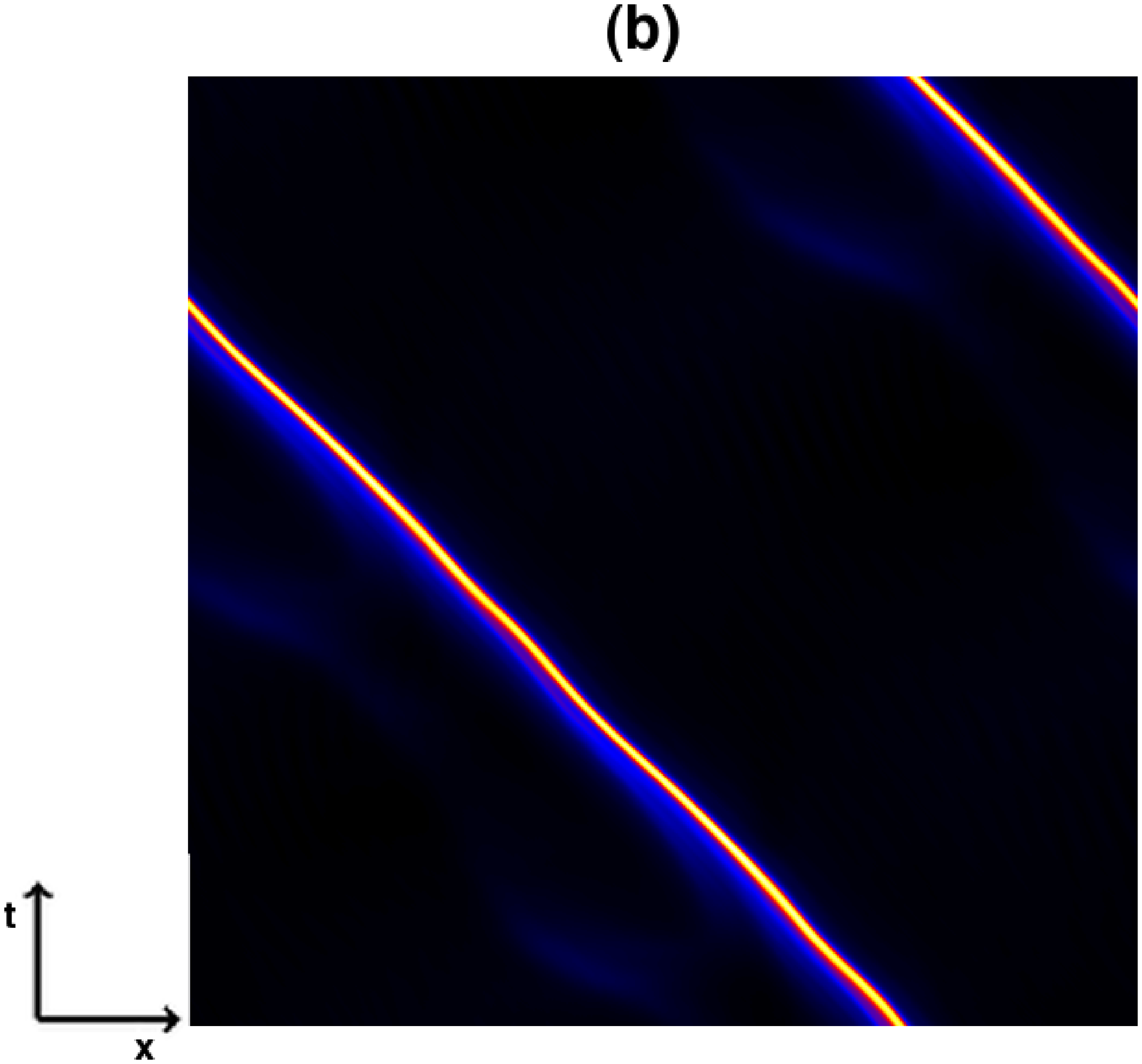}
\end{center}
\caption{(Color online) (a) Component $\sigma_{11}$ of the conformation tensor.
(b) Space-time plot obtained by following the evolution in time of a horizontal
cross section (corresponding to the dotted line in (a)) of
$\sigma_{11}(x,y,t)$.  Time is plotted versus position; lines of constant slope
mean the pattern moves at constant velocity.  Due to periodic boundary
conditions, waves leaving the box on the left re-enter on the right. Here
$Re=0.875$, $Wi=14$.}
\label{fig:fig08}
\end{figure}

Since patterns described so far travel along the longitudinal direction, it is
natural to wonder what their typical speed is, how this possibly varies with
the transversal coordinate, and how these quantities change when the elasticity
of the solution is increased.  We measured the speed of these elastic waves by
tracking peaks of intensity in the fields $\zeta$, $\sigma_{ij}$ ($i,j=1,2$),
and following their motion along the $x-$direction (Fig.~\ref{fig:fig08}).

\begin{table}[!h]
\begin{center}
\begin{tabular}{|c|c|c|c|c|c|c|}
\hline
$Re_0$ & $Wi_0$ & $Re$ & $Wi$ & $U$ & $\langle |v| \rangle \pm \delta |v|$ & $\langle|v|\rangle/U$\\
\hline
\hline
$1$ & $12$ & $0.925$ & $11.1$ & $3.7$ & $1.938 \pm 0.016$ & $0.52$ \\
\hline
$1$ & $16$ & $0.875$ & $14.0$ & $3.5$ & $1.90 \pm 0.12$ & $0.54$ \\
\hline
$1$ & $20$ & $0.813$ & $16.3$ & $3.25$ & $1.70 \pm 0.27$ & $0.52$ \\
\hline
$0.1$ & $12$ & $0.093$ & $11.1$ & $3.7$ & $1.704 \pm 0.022$ & $0.46$ \\
\hline
$0.05$ & $12$ & $0.049$ & $11.7$ & $3.9$ & $1.722 \pm 0.032$ & $0.44$ \\
\hline
\end{tabular}
\end{center}
\caption{Measure of the typical speed of waves for various values of the
Reynolds and Weissenberg numbers.  Here $Re_0$ and $Wi_0$ are computed from the
laminar fixed point velocity amplitude $U_0=4$; $U$ is the measured intensity of the mean
velocity profile, from which $Re$ and $Wi$ are estimated.}
\label{tab:velonde}
\end{table}

\begin{figure}[!h]
\includegraphics[draft=false,scale=0.55]{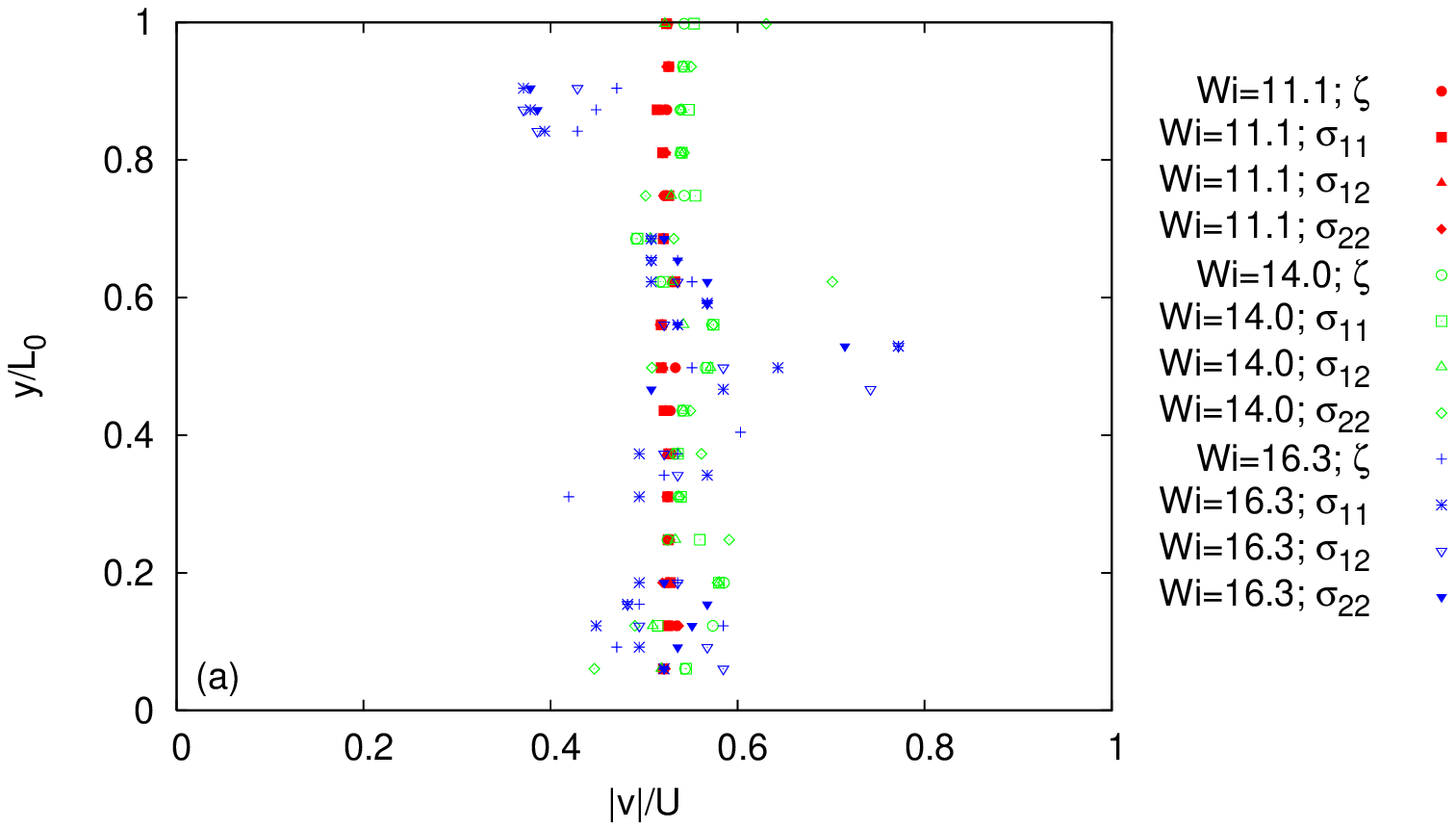}
\includegraphics[draft=false,scale=0.55]{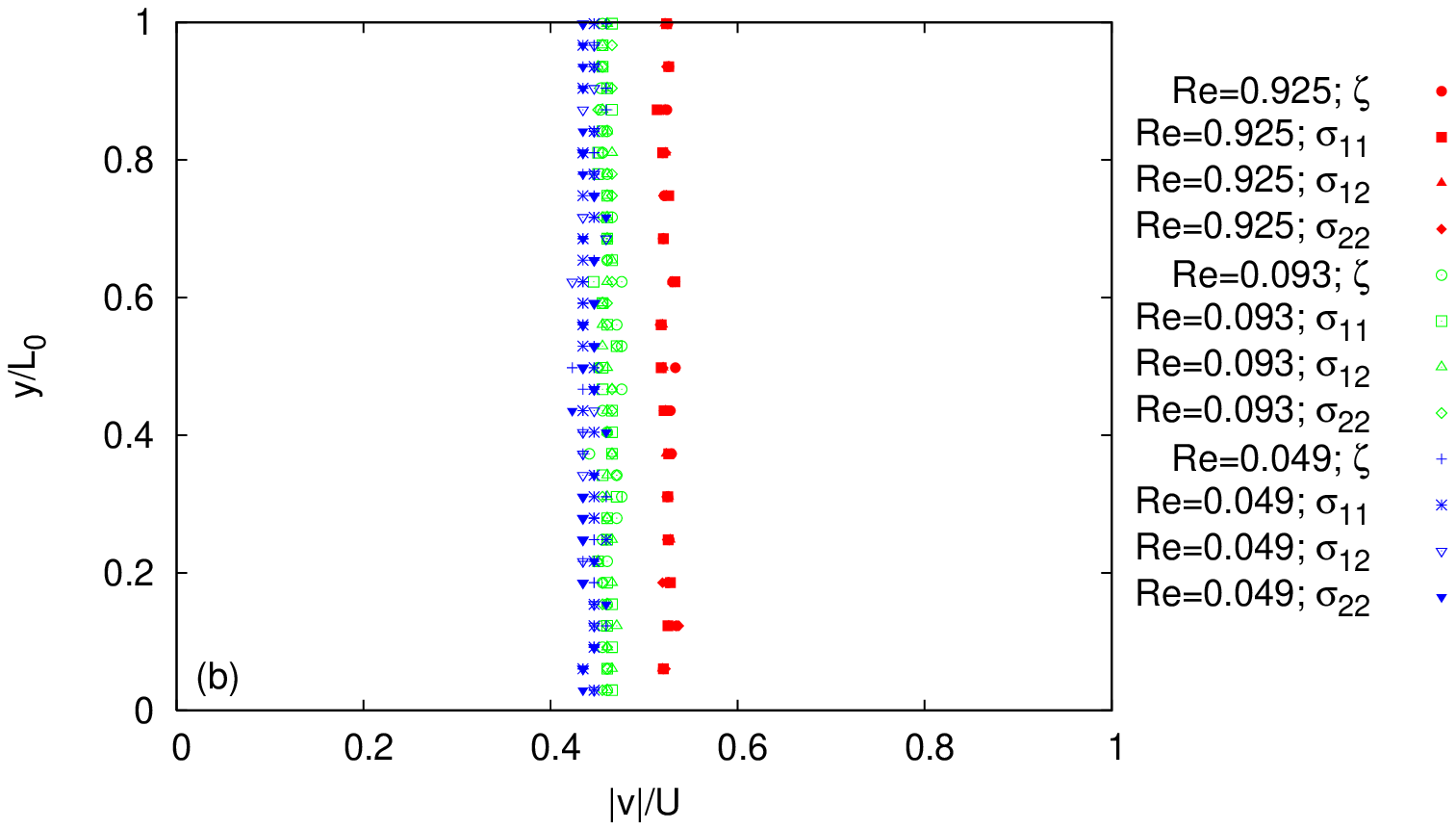}
\caption{(Color online) Transversal profiles of wave speed: (a) measurements at
fixed $Re \simeq 1$ and varying Weissenberg number $Wi=11.1$ (red), $Wi=14$
(green), $Wi=16.3$ (blue); (b) measurements at fixed $Wi \simeq 11.1$ and
varying Reynolds number $Re=0.925$ (red), $Re=0.093$ (green), $Re=0.049$
(blue). In both plots wave speeds measured in  different
fields ($\zeta$, $\sigma_{11}$, $\sigma_{12}$
$\sigma_{22}$) are presented. Wave speeds are made non-dimensional dividing 
by the intensity of the mean velocity profile $U$, the vertical coordinate is 
normalized to the box size $L_0$.}
\label{fig:fig09}
\end{figure}

The results relative to the cases where coherent motion of structures was
detected are reported in Table~\ref{tab:velonde} and Fig.~\ref{fig:fig09}.  In
all the considered cases we found an average wave speed intensity close to half
the peak fluid velocity: $\langle |v| \rangle \simeq U/2$.  Concerning spatial
variations of $|v|$, these show little dispersion $\delta |v|$ around the mean
at $Wi=11.1$, as it can be seen also in the plot (Fig.~\ref{fig:fig09}) of the
profiles along the transversal coordinate $y$, which are essentially constant.
The speed of different parts of a wave, for instance its tip and its tail,
tracked by cross-sections at different $y$ coordinates, move at approximately
the same speed, pointing to rigid motion of the structure, which travels
substantially undeformed. We remark these dynamics are compatible with the
constancy of the global quantities shown in Fig.~\ref{fig:fig03}, due to the
fact that $K$ and $\Sigma$ are integrals over the space domain by definition,
and to the presence of periodic boundary conditions. 
When the Weissenberg number is increased ($Wi=14$ in Fig.~\ref{fig:fig09}a),
wave speed profiles become less constant, indicating more complex motion. The
larger scatter of $v$ values reflects the larger uncertainty $\delta |v|$ in
the measure of the propagation speed of patterns, which are now found to swing
at $y$-locations corresponding to their tails, and the presence of secondary
less steady filamental structures in regions between the better defined primary
waves.  At even larger values of the Weissenberg number ($Wi=16.3$), waves
interact considerably more, starting to move in both directions and to break
and reform after they ``collide''; the distribution of wave speeds now exhibits
significantly more variability along the $y-$direction.  In fact, at the
present Reynolds number ($Re \simeq 1$), this is the largest $Wi$ at which we
succeeded in tracking waves. Above this value, patterns form more easily in
various parts of the spatial domain, but interactions among them dominate the
dynamics, producing a disordered flow in both space and time.

At the smallest value of $Wi=11.1$ we find that the dependence of the mean
velocity on $Re$ is very weak (see Fig.~\ref{fig:fig09}b), confirming 
the elastic nature of these instabilities.

\begin{figure}[!h]
\begin{center}
\includegraphics[draft=false,scale=0.155]{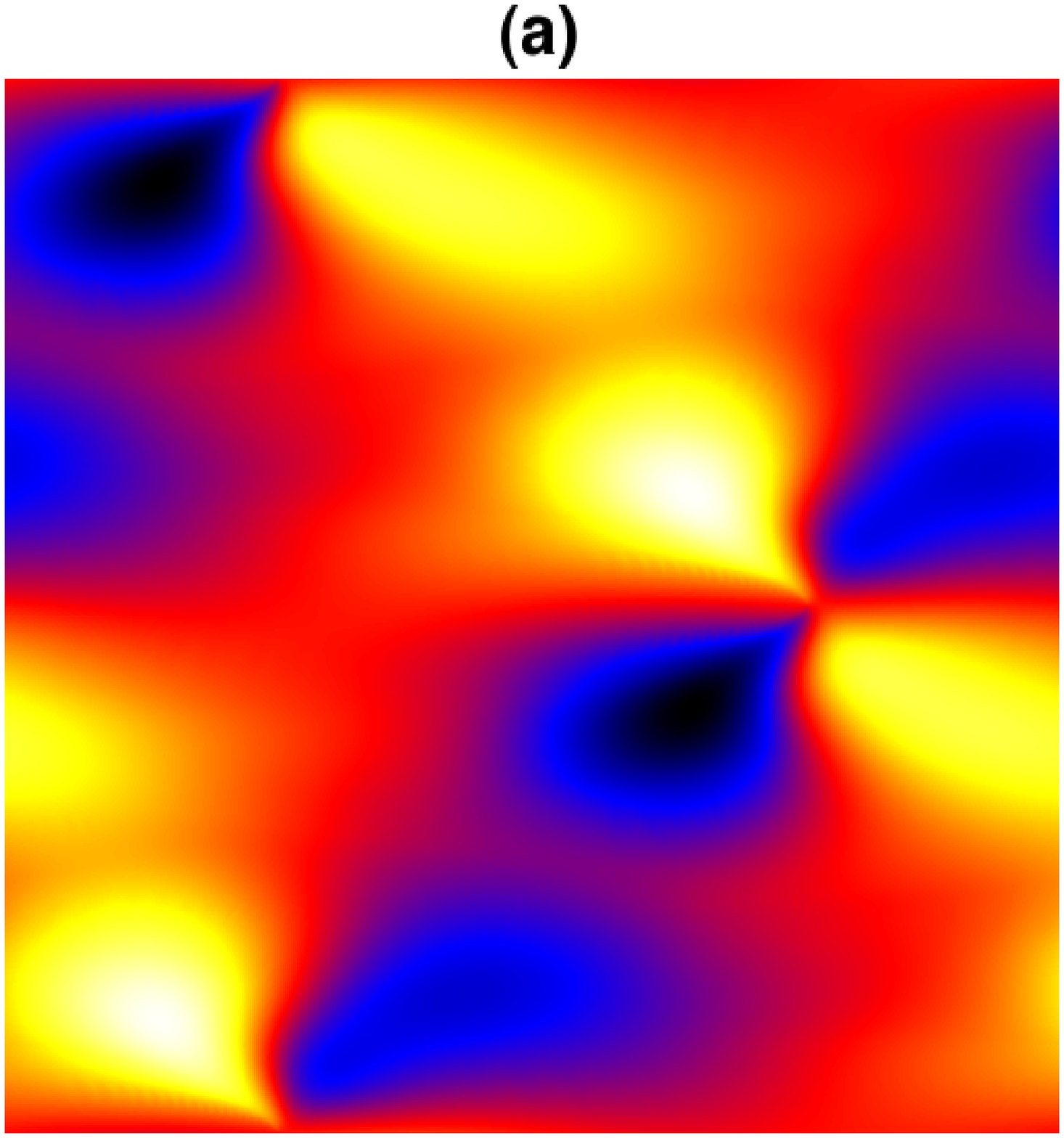} 
\includegraphics[draft=false,scale=0.155]{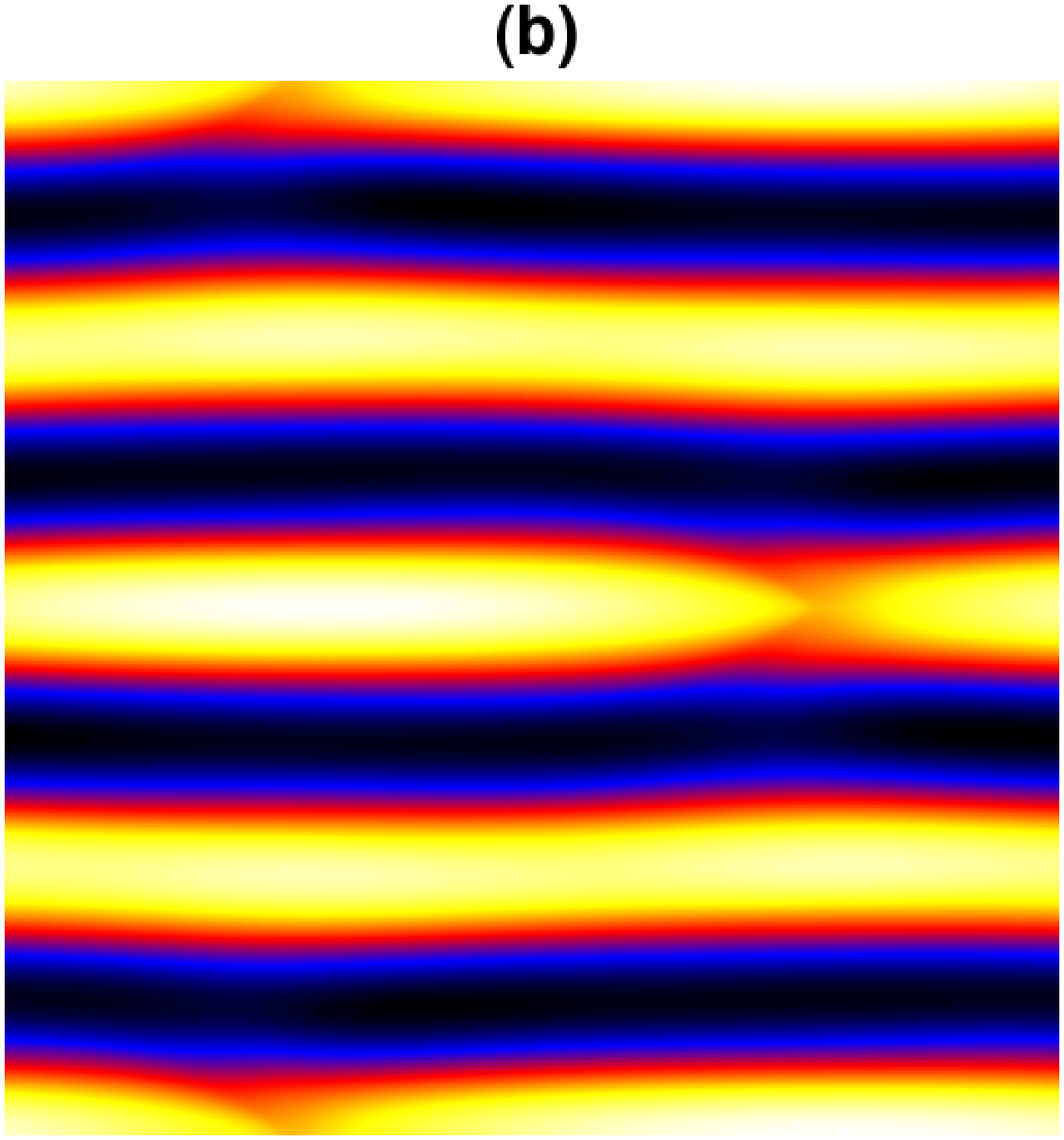}
\includegraphics[draft=false,scale=0.155]{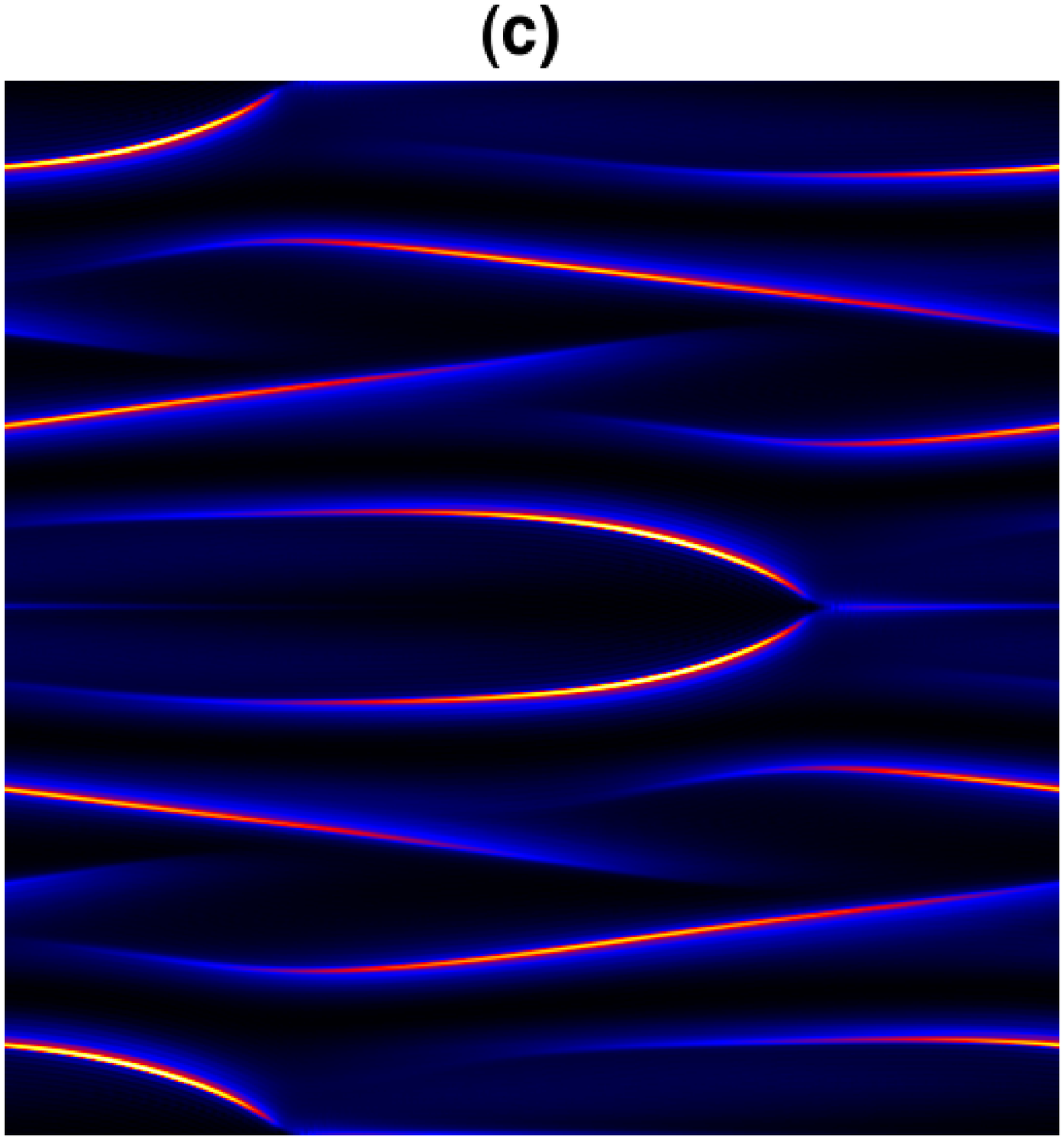}
\end{center}
\caption{(Color online) Snapshots in the $(x,y)$~plane of different fields at
$Re=0.925$, $Wi=11.1$.  (a) Secondary transverse flow $u_y \equiv u_y'$; (b)
longitudinal flow $u_x=u_{0x}+u_x'$; (c) square polymer elongation $tr
\bm{\sigma}$.  In all cases black means minimum and white 
maximum intensity.  Due to the positive-definiteness of the conformation
tensor,  in (c) black corresponds to zero average square elongation.}
\label{fig:fig10}
\end{figure}

In order to get some insight into the physical mechanism producing filamental
patterns, we now consider the spatial structure of the components $u_x$, $u_y$
of the velocity field, for $Re \simeq 1$ and $Wi \gtrsim Wi_c$. The velocity
field can be thought as the sum of the base flow and a secondary flow:
$u_x=u_{0x}+u_x'$, $u_y=u_{0y}+u_y'$; at the laminar fixed point, of course
$u_x=u_{0x}=U_0\cos(y/L)$, $u_y=u_{0y}=0$. Here we are especially concerned
with the establishment of a nonzero transversal velocity component and the
breaking of translational invariance along the longitudinal direction, above
the elastic instability threshold.  Snapshots at fixed time of the fields
$u_x(x,y,t)$ and $u_y(x,y,t)$, computed from the vorticity field $\zeta$ shown
in Fig.~\ref{fig:fig07}a, are presented in
Fig.~\ref{fig:fig10}a,~\ref{fig:fig10}b. 

In the horizontal component of velocity $u_x$, the periodicity of the base flow
is clearly visible. The transversal component $u_y$, on the other hand,
displays localized quadrupolar patterns of positive and negative values, on a
background of zero velocity.  Just to fix the ideas, consider, for instance,
the pattern centered around $y=L_0/2$, $L_0=2\pi$ being the height of the
computational domain, on the right part of Fig.~\ref{fig:fig10}a. Notice that
the position of this pattern matches exactly that of the tip of the most
clearly visible wave in the vorticity field (Fig.~\ref{fig:fig07}a). 
Indeed, this structure is found to be stable and to travel along the horizontal
direction, following the propagation of the vorticity filament.  The lobes of
these quadrupolar patterns (with white, black indicating positive, negative values
respectively), appear in correspondence to locations of strong primary strain,
where gradients of the base flow attain maximal values in magnitude.  In these
regions polymers are significantly stretched, by the terms $(\bm{\nabla}
\bm{u})^T \cdot \bm{\sigma}$ and $\bm{\sigma} \cdot (\bm{\nabla} \bm{u})$ in
eq.~(\ref{eq:eq02}). In Fig.~\ref{fig:fig10}c a snapshot, at the same time,
of the trace of the conformation tensor, which measures the square elongation
of polymers, is reported. 
Strong elongations appear in thin filamental structures localized around $y$
coordinates where $u_x$ changes sign or, equivalently, where $\partial_y u_x$
is maximum.  As a consequence, it is in this regions that they can exert a
relevant feedback on the flow, through the elastic stress term ${{2\eta\nu}
\over \tau} \bm{\nabla} \cdot \bm{\sigma}$.  

\begin{figure}[!h]
\includegraphics[draft=false,scale=1]{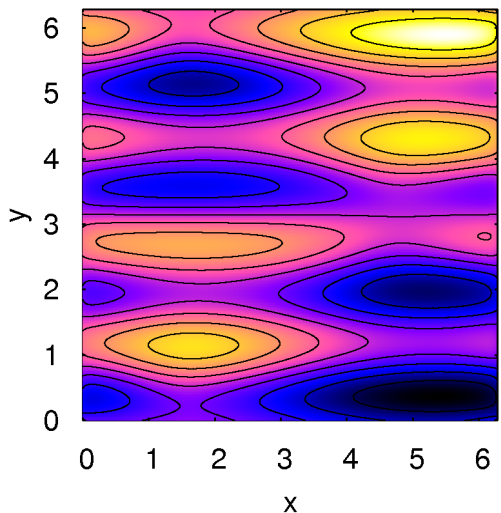}
\caption{(Color online) Stream function $\psi(x,y)$ for $Re=0.925$, 
$Wi=11.1$; black corresponds to minimum, white to maximum.}
\label{fig:fig11}
\end{figure}

In other words, the transverse flow develops where the feedback of polymers on
the velocity dynamics is strongest. Finally, the establishment of the
transversal component of the flow is accompanied by a reduction of the
longitudinal one, due to incompressibility (Fig.~\ref{fig:fig10}b). The result
of this whole process is the formation of recirculation zones, which are
evident in a plot of the corresponding stream function $\psi$ (such that
$u_x=-\partial_y \psi$, $u_y=\partial_x \psi$) (Fig.~\ref{fig:fig11}). These in
turn result to be organized in an array of counter-rotating vortices of equal
strength, with a basic periodicity of $2$ in the longitudinal direction, which
propagates horizontally. 

\begin{figure}[!h]
\includegraphics[draft=false,scale=0.6]{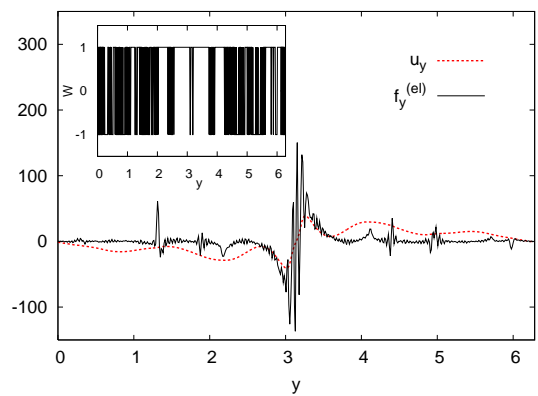}
\caption{(Color online) Transverse polymer body force $f^{(el)}_y$ and
transverse velocity $u_y$ (multiplied by a factor $150$ for reasons
of visualization) versus the vertical coordinate $y$. Inset: elastic force's
power as a function of the vertical coordinate.  Here $Re=0.925$, $Wi=11.1$ and
the horizontal coordinate is $x \simeq 4.7$.  This corresponds to a vertical
line immediately on the left of the center of the  quadrupolar pattern in
Fig.~\ref{fig:fig10}.}
\label{fig:fig12}
\end{figure}

Concerning the feedback of polymers on the flow field, the transversal velocity
is found to be sustained by a persistent correlation between the transversal
components of the velocity field $u_y$ and the elastic body force $f^{(el)}_y=
\left( 2 \eta \nu/\tau \right) \left( \bm{\nabla}  \cdot \bm{\sigma}
\right)_y$, confirming what anticipated in the preceding discussion. The
$y-$profiles of these quantities are shown in Fig.~\ref{fig:fig12}, at
$x\simeq4.7$, that is immediately on the left of the center of the quadrupolar
pattern discussed above.  Though this figure refers to a single realization at
fixed time, this positive correlation is a feature persisting over time.  In
Fig.~\ref{fig:fig12} it is also shown the profile of $W=f^{(el)}_y u_y
/(|f^{(el)}_y |\cdot|u_y|)$, which is the sign of the elastic force's power.
Though this fluctuates wildly, it is clearly positive in finite size regions
located around $y\simeq\pi$, corresponding to the extension of the quadrupolar
pattern on the right in fig~\ref{fig:fig10}a. This observation further supports
the role of elastic contributions in the establishment of the transversal flow,
and hence in the appearance of filamental structures.

\begin{figure}[!h]
\includegraphics[draft=false,scale=0.3]{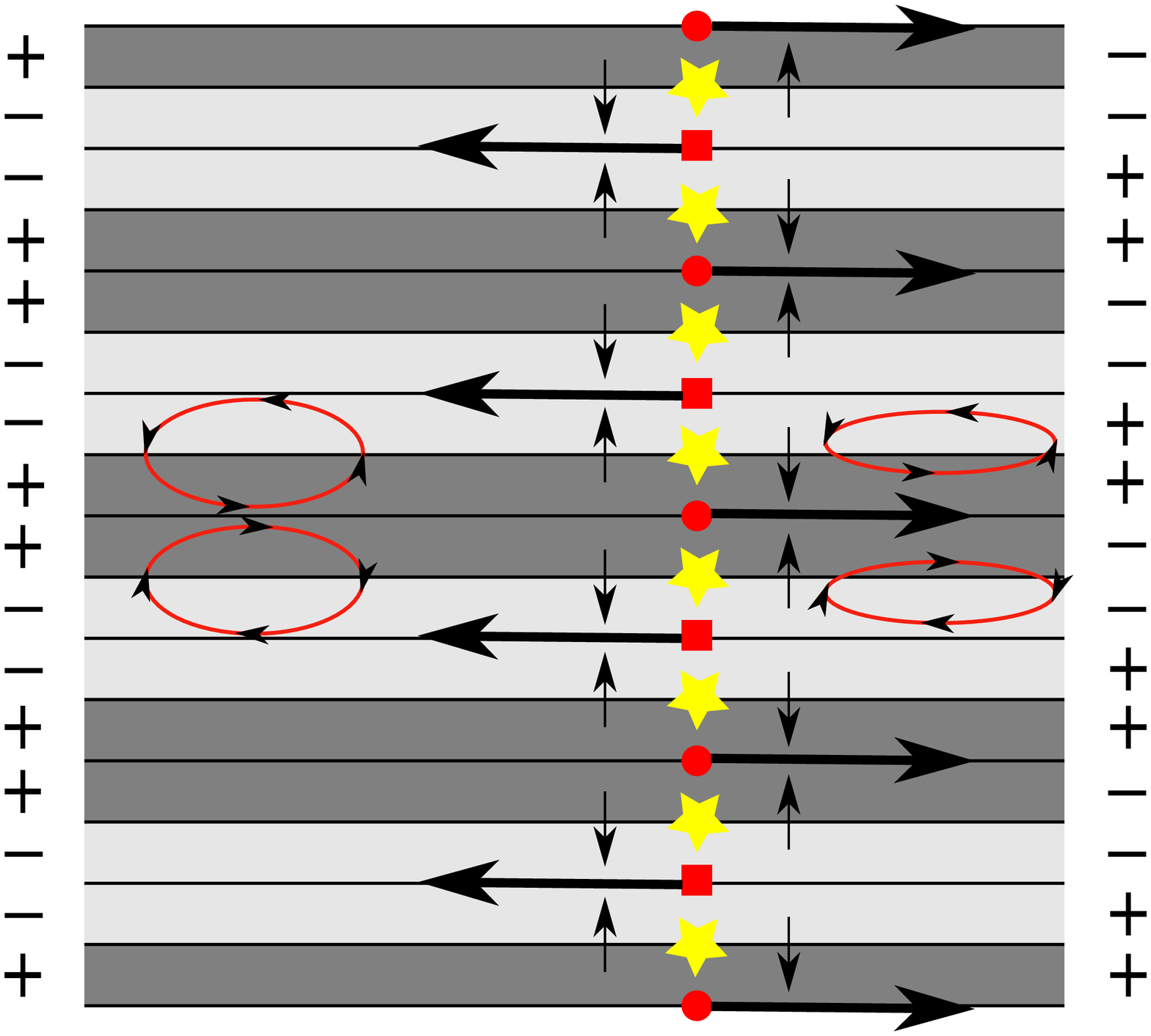}
\caption{(Color online) Schematic illustration of the formation of vortices.
Yellow star points indicate locations of maximum velocity gradient in absolute
value, where polymers are significantly stretched and the transverse flow is
more likely to occur; small black arrows correspond to the secondary
transversal flow; big black arrows give the direction of propagation of waves.
The left column contains the sign of $u_{0x}$ and the right one that of
$\zeta$.}
\label{fig:fig13}
\end{figure}

\begin{figure}[!h]
\includegraphics[draft=false,scale=0.6]{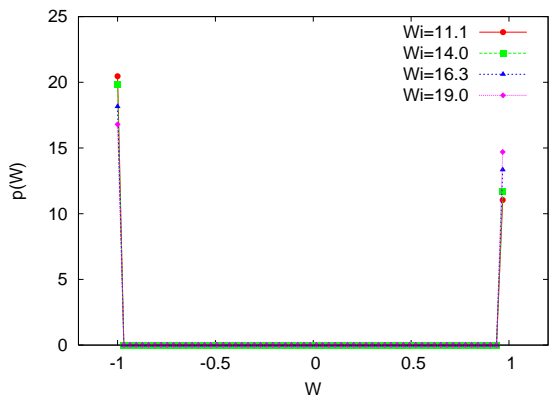}
\caption{(Color online) Probability density function of the sign of transverse
flow's power.  At growing $Wi$ positive values are more likely to occur. Here
the Reynolds number is always $Re \simeq 1$.} \label{fig:fig14}
\end{figure}

To conclude, the picture emerging from this description is that filamental
patterns are associated to the formation of arrays of counter-rotating
vortices, sustained by the dynamical coupling between velocity and localized
structures where polymers are strongly elongated. Indeed this is probably the
simplest way to break the translational symmetry of the velocity field along
the horizontal direction. The heuristic argument developed here is summarized
in the scheme in Fig.~\ref{fig:fig13}. As a final remark, it may be interesting
to observe that this allows to understand the direction of wave propagation.
This results from the direction of the base velocity field, from which vortices
of intensity of order $U/2$ form.  Due to the periodicity of the base flow,
both positive and negative horizontal directions of propagation are possible.
The selected direction then only depends on where the positive correlation
between a random perturbation of $u_y$ and the elastic body force $f_y^{(el)}$
happens to be established. As shown in Fig.~\ref{fig:fig14}, at increasing the
elasticity of the solution, the probability to find positive correlations gets
larger, which implies the possibility of forming filaments in a larger fraction
of the domain. This reflects the experimental observation that when $Wi$ is
sufficiently large, filaments' propagation may happen in both directions.

\section{Summary and conclusions}
\label{sec:sec04}
In this work we have studied the dynamics of a two-dimensional parallel flow,
namely the Kolmogorov shear flow, of a viscoelastic fluid, close to the elastic
instability threshold predicted in~\cite{BCMPV05} within the framework of
linear stability analysis.  By means of direct numerical simulations of the
Oldroyd-B model, we investigated the destabilization of this flow induced by
the elastic forces associated to the dynamics of polymer molecules in the
solution.  Above a critical Weissenberg number $Wi_c \approx 10$ a transition
to new dynamical states was observed.  In this regime of large elasticity, the
system was found to be characterized by the presence of filamental structures
propagating in the direction of the mean flow and global oscillations
introducing temporal dependencies in the observables. This phenomenon allows
the establishment of mixing features. Close to the critical Weissenberg number
the structures can be described as traveling waves associated to steady
patterns of counter-rotating vortices sustained by the coupling between elastic
forces and velocity field. The onset of wavy motion of filaments was found to
occur even in the limit of very small Reynolds numbers, typically for $Re \geq
0.05$, at $Wi \gtrsim Wi_c$.

Increasing the Weissenberg number, the system evolves towards more complex
dynamics, where filaments are observed to continuously form and break as a
consequence of interactions among them. As a consequence, several time scales
and finer spatial scales result to be involved in the dynamical behavior,
which displays strongly non-linear features resembling those of a turbulent
flow. For instance, when $Wi \gtrsim 20$ (and $Re\lesssim 1$) the power
spectrum of velocity fluctuations $E(k)$ is given by a power-law $k^{-\alpha}$
of exponent $\alpha$ larger than $3$, in close agreement with experimental
findings~\cite{GS00}.

The present results refer to a simplified viscoelastic model in an idealized
geometrical configuration.  Nevertheless, this simple setup proved useful to
detect the appearance of coherent structures in the form of ``elastic waves'',
driving the transition towards mixing states, and to reproduce the basic
phenomenology observed in experiments.  We hope these findings may help in the
understanding of the phenomenon  of elastic turbulence in more realistic
configurations and in the comprehension of the elastic instability mechanism in
parallel flows.

\section*{ACKNOWLEDGMENTS}
We are grateful to S.~Rafa\"i for interesting discussions and a critical
reading of the manuscript. S.B. acknowledges financial support from CNRS.

\end{document}